# High Spatial Resolution Fast-Neutron Imaging Detectors for Pulsed Fast-Neutron Transmission Spectroscopy

**Mor I.**[a*], **Vartsky D.**[a], **Bar D.**[a], **Feldman G.**[a], **Goldberg M. B.**[a†], **Katz D.**[a], **Sayag E.**[a], **Shmueli I.**[a], **Cohen Y.**[a], **Tal A.**[a], **Vagish Z.**[a], **Bromberger B.**[b], **Dangendorf V.**[b], **Mugai D.**[b], **Tittelmeier K.**[b], **Weierganz M.**[b]

[a] *Soreq NRC,*
*Yavne 81800, Israel*
[b] *Physikalisch-Technische Bundesanstalt (PTB),*
*38116 Braunschweig, Germany*
*E-mail*: ilmor@soreq.gov.il

ABSTRACT: Two generations of a novel detector for high-resolution transmission imaging and spectrometry of fast-neutrons are presented. These devices are based on a hydrogenous fiber scintillator screen and single- or multiple-gated intensified camera systems (ICCD). This detector is designed for energy-selective neutron radiography with nanosecond-pulsed broad-energy (1 - 10 MeV) neutron beams. Utilizing the Time-of-Flight (TOF) method, such a detector is capable of simultaneously capturing several images, each at a different neutron energy (TOF). In addition, a gamma-ray image can also be simultaneously registered, allowing combined neutron/gamma inspection of objects. This permits combining the sensitivity of the fast-neutron resonance method to low-Z elements with that of gamma radiography to high-Z materials.

* Corresponding author.
† On Sabbatical leave at PTB-Braunschweig

## Contents

# 1. Introduction

Detection of explosives concealed in air-cargo or passenger-baggage presents a considerable challenge which has not been fully met by currently-deployed X-ray and gamma-ray inspection systems. These systems provide only limited information about the objects contained, such as their shape and density, and their performance capabilities rely heavily on human operator skill. Furthermore, only very limited differentiation amongst elements in the low atomic-number (Z) range can be achieved. The sole x-ray based explosives detection technique capable of automatic detection utilizes coherent x-ray scattering [1]. However, its penetration through items that are substantially more massive than carry-on baggage is severely limited by the low energy of the relevant X-rays (<100 keV).

Fast-Neutron Resonance Radiography (FNRR) is one of the most promising methods for fully-automatic detection and identification of explosives concealed in passenger luggage and air-cargo. Neutron transmission depends only weakly on absorber Z **(**NOTA BENE: with the exception of hydrogen**)**, allowing neutrons to penetrate voluminous objects and high-Z materials. In addition, neutrons probe the nuclear properties of the absorber and exhibit highly characteristic structure in the neutron-energy dependence of interaction cross-sections with different isotopes.

Over the years, various neutron-based imaging methods were considered for use in screening baggage and air- cargo [2-7]. These methods enable the determination of the volume distributions of H, C, N, O, and possibly other elements contained within transported substances. One of the neutron-based methods proven of revealing the elemental composition of materials is Pulsed Fast Neutron Transmission Spectroscopy (PFNTS). This method is based on measurement of the neutron transmission through the inspected object as function of neutron energy. The energy spectrum of the transmitted neutrons is modified according to cross-section fluctuations ("peaks" and "dips") exhibited by light elements in the 1-10 MeV energy range.

Fig. 1 shows the calculated transmission spectra of MeV neutrons through 10 cm material of polyethylene, Tri-Acetone Tri-Peroxide (TATP[1]) and melamine. The various peaks and dips in each spectrum are characteristic of its elemental composition, and specifically, of its relative abundance of carbon, oxygen and nitrogen.

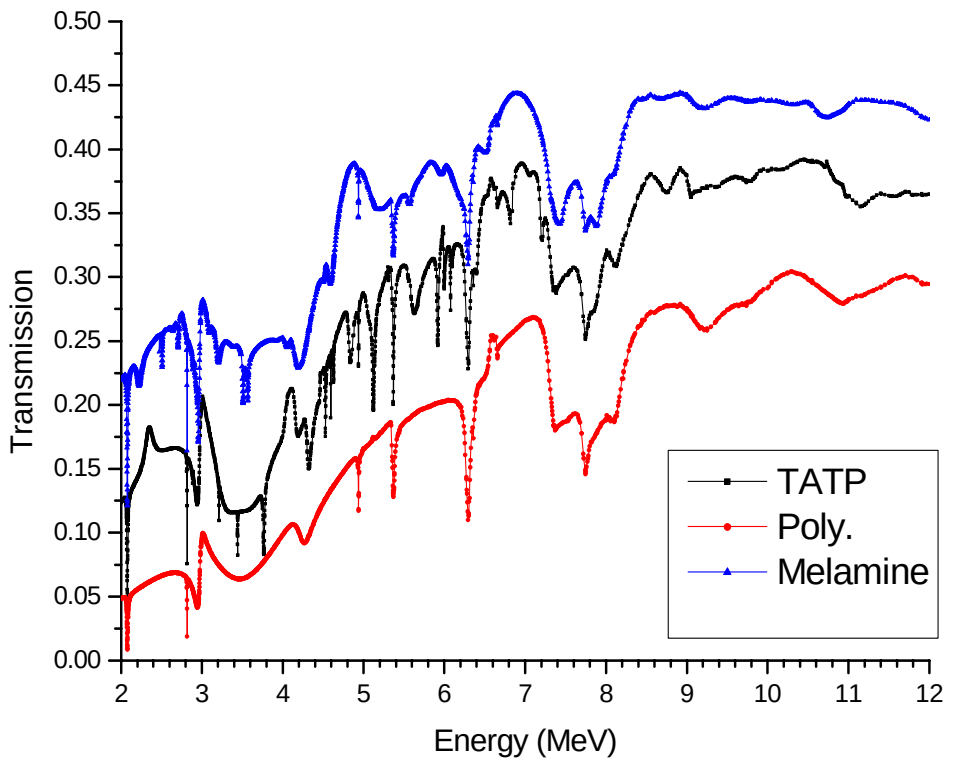

**Fig. 1 Calculated transmission through 10 cm of TATP ($C_9H_{18}O_6$), polyethylene $(C_2H_4)_n$ and melamine ($C_3H_6N_6$)**

Using the time-of-flight (TOF) technique, the spectrum of the transmitted neutrons is measured by a position-sensitive neutron detector and the attenuation at pre-selected time-intervals (neutron energies) is determined.

The PFNTS method was proposed and first studied by Oregon University as of 1985 [2, 3, 8-10] and subsequently refined by Tensor-Technology Inc. [4, 11-15], culminating in series of blind tests conducted during the late 1990s by the US Federal Aviation Administration (FAA) [4]. The results of these tests demonstrated that PFNTS constitutes a powerful detection method for detecting bulk explosives.

[1] TATP is an improvised explosive made of standard household chemicals, that has been employed by terrorists in Israel, the U.K. and a recently foiled attempt in Germany

However, the pixel size defined by these detectors (several centimeters) imposed an intrinsic limitation on the spatial resolution, which prohibited reliable detection of objects smaller and thinner than, typically, 2-3 cm. Thus, in 1999, the National Materials Advisory Board Panel [4], advised against constructing an operational airport prototype, since no compact, suitable neutron source was available at that time, nor was the detector spatial resolution adequate for reliable detection of thin-sheet explosives. Nevertheless, the panel did conclude that a mature inspection system of this type, being completely automatic, would represent a significant improvement over systems available then (and now).

In order to respond to the above challenge, a next-generation PFNTS detector should be capable of providing mm-size spatial resolution and good TOF spectroscopy (few ns which translates to several hundred keV) per pixel as well as the ability to operate at high neutron fluxes ($\geq 10^6$ n/(s×cm$^2$)). The following describes two generations of TRION (Time Resolved Integrative Optical (readout for) Neutrons) detectors that attempt to meet the above requirements. These represent significant stages of development towards a detector deployable in an operational PFNTS inspection system.

## 2. The TRION detector

### 2.1 Concept and results with first generation single-frame system

TRION is a novel, fast-neutron imaging device based on time-gated optical readout. The concept was first proposed by Dangendorf et al. [16] of PTB and subsequently developed at Soreq NRC and PTB. The first generation detector, developed and evaluated in several beam experiments, was capable of imaging only a single TOF-frame per beam burst. The basic design of the single-frame TRION detector is shown in Figs. 2 and 3 and will be referred to in the following as TRION Gen.1.

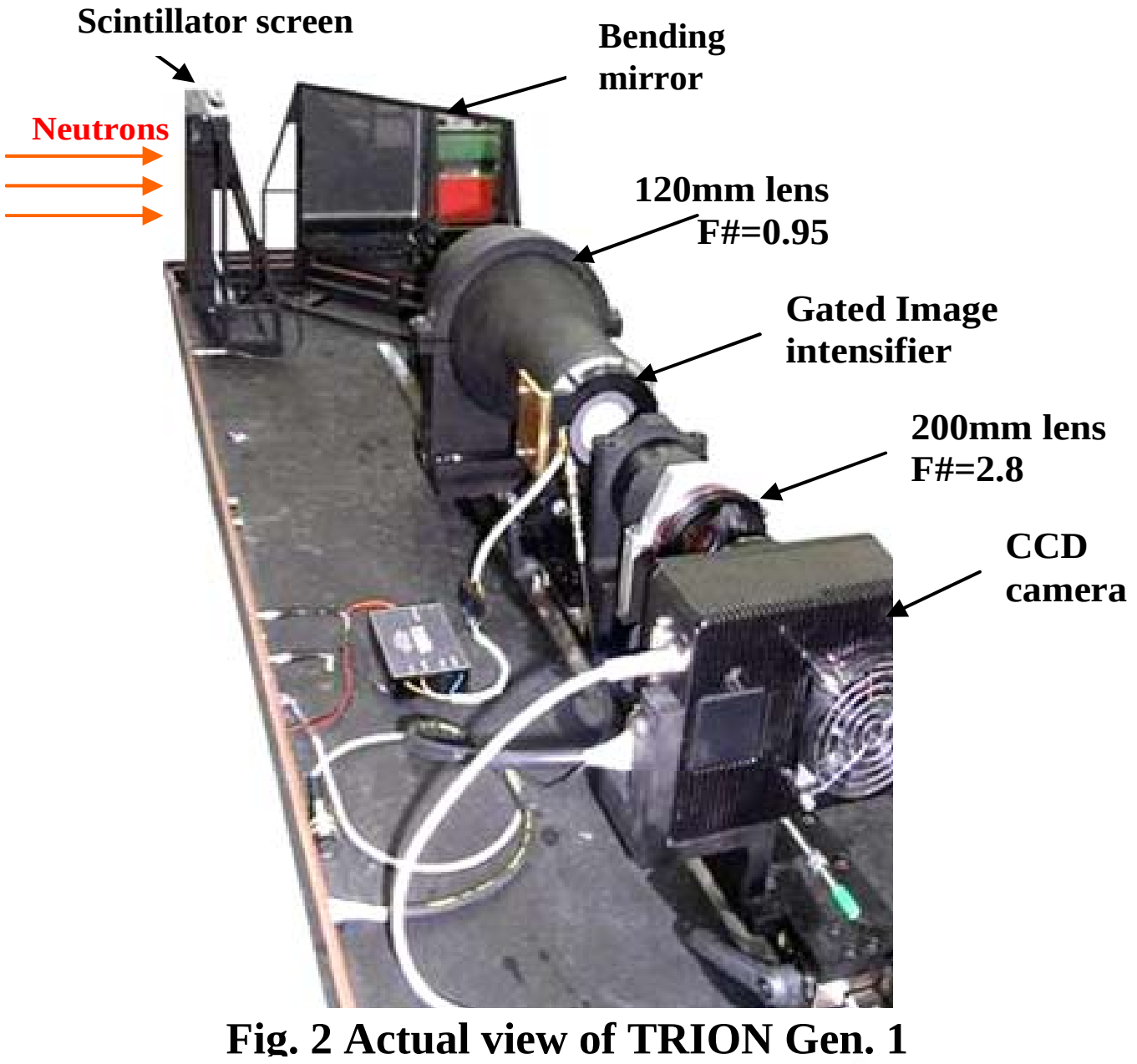

**Fig. 2 Actual view of TRION Gen. 1**

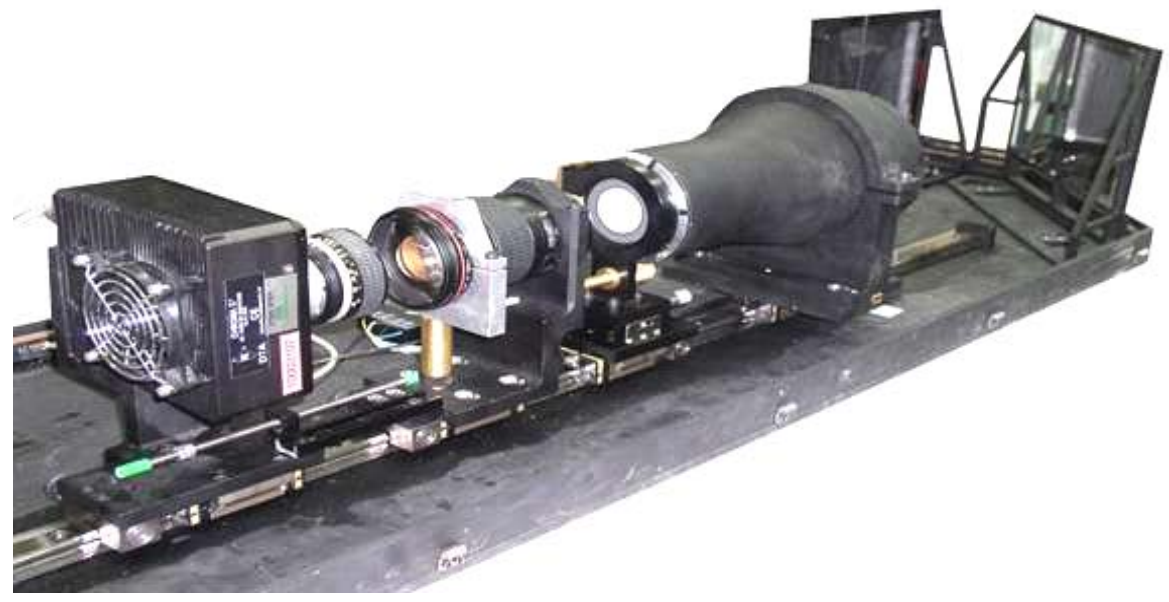

**Fig. 3 Side view of TRION Gen. 1**

The detector is designed to detect fast-neutron pulses produced, for example, in the $^9$Be(d,n) reaction using a pulsed (~1-2 ns bursts, 1-2 MHz repetition rate) deuteron beam. After a specific Time-of-Flight (TOF) that depends on their energy, the fast-neutrons impinge on the plastic fiber scintillator, causing the emission of light from the screen via knock-on protons. The light is reflected by a front-coated bending mirror (98% reflectivity), positioned at an angle of $45^{\circ}$ relative to the neutron beam direction, towards a large aperture 120 mm F#0.95 lens and subsequently focused on the image-intensifier. The latter not only amplifies the light intensity but, more importantly, acts as an electronic shutter that is opened for a gate period of $\Delta t$ (as short as 8 ns) at a fixed, pre-selected TOF relative to each beam burst. Repetition rate for the beam pulses was up to 2 MHz and images were integrated by cooled CCD camera over many beam bursts, with acquisition times ranging from tens to several hundreds of seconds.

The system components are mounted on linear guides that can move freely along a precision rail. All system components are mounted in a light-tight enclosure.

### 2.1.1 Components

The following provides a more detailed description of each TRION component.

#### *2.1.1.1 Scintillator screen*

A plastic fiber scintillator screen, shown in Fig. 4, is used in this work as neutron detector. The most common fast-neutron detection scheme is based on conversion of the neutron into a proton, via neutron elastic-scattering by hydrogen present in the detector medium.

The fiber screen, BCF 99-55 (based on BCF-12 fiber [17]), was manufactured by Saint-Gobain (formerly Bicron), USA. Its light emission peaks at 435 nm (see Fig. 4 [17]) with decay time of 3.2 ns. The screen surface area is 200×200 $mm^2$ and is made of 30 mm long scintillating fibers, each consisting of a 0.5×0.5 $mm^2$ polystyrene core (refractive index-$n_1$=1.60), a 20 μm thick Poly-Methyl-MethAcrylate (PMMA) cladding (refractive index-$n_2$=1.49), and a 16 μm thick, $TiO_2$ doped, white polyurethane paint coating, which acts as Extra Mural Absorber (EMA) to prevent light cross-talk. The fibers were first assembled in 10×10 $mm^2$ square bundles, which were then glued together to form the entire screen. The face of the fiber screen facing the incident neutron beam is covered by a backing mirror (of 98% reflectivity, made by Praezisions Glas &

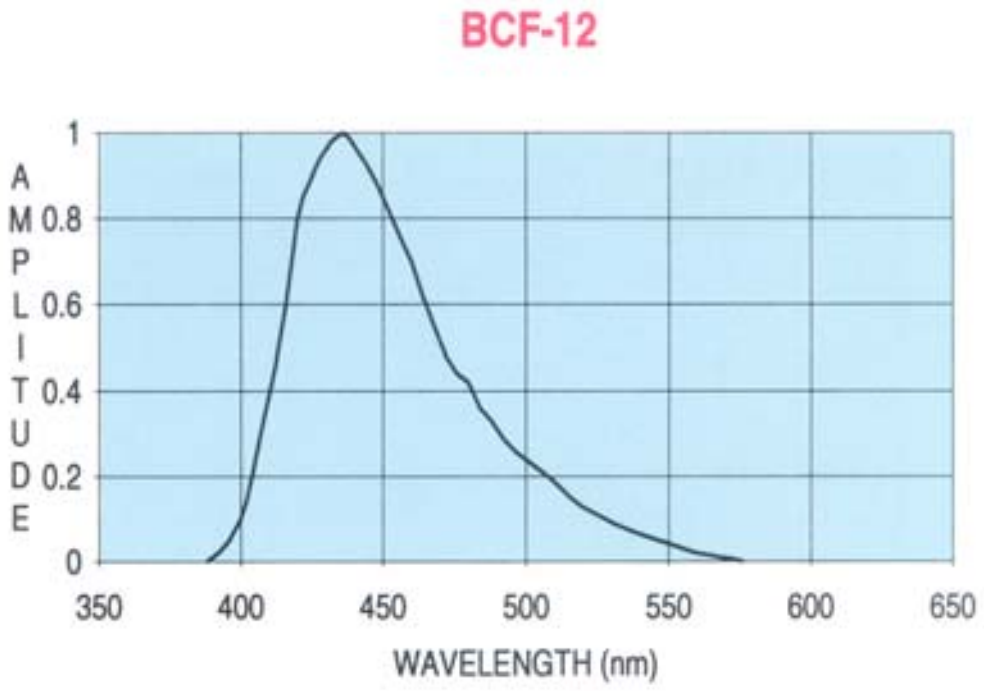

**Fig. 4 BCF 12 emission spectrum**

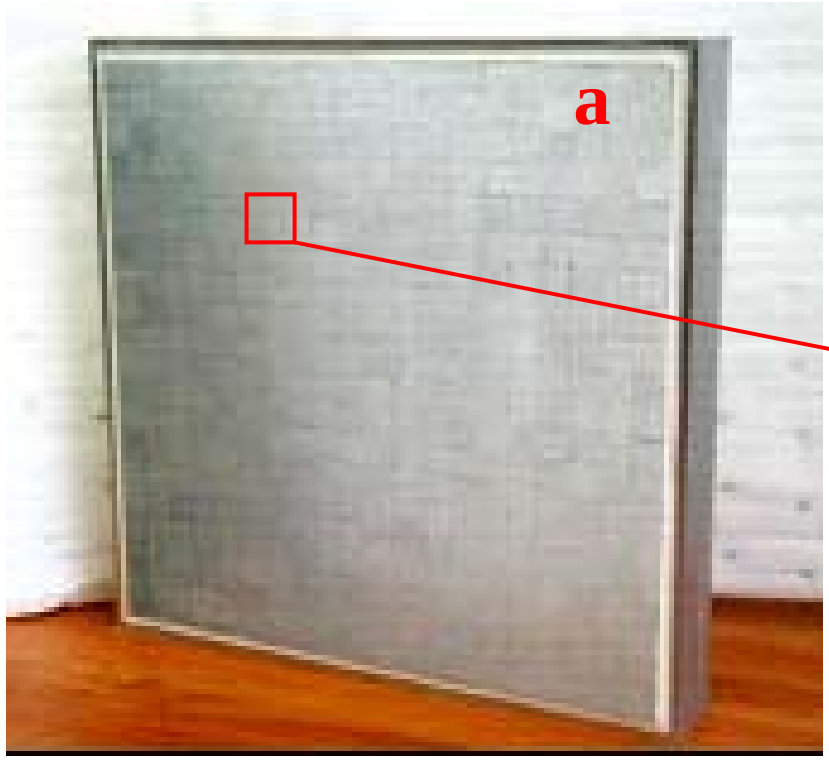

**Fig. 5a Fiber scintillator screen used in TRION**

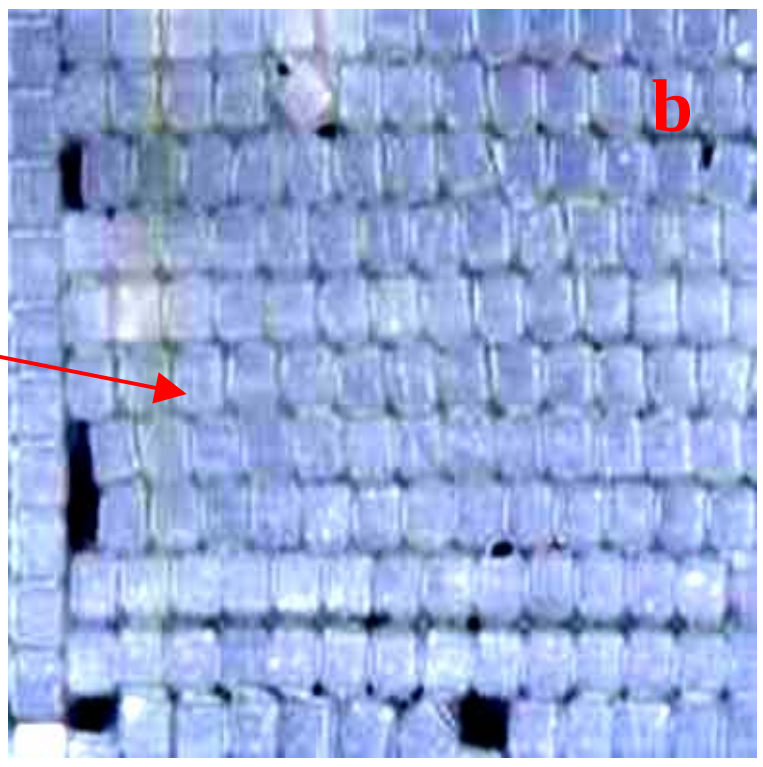

**Fig. 5b Enlarged photograph of a section from the screen**

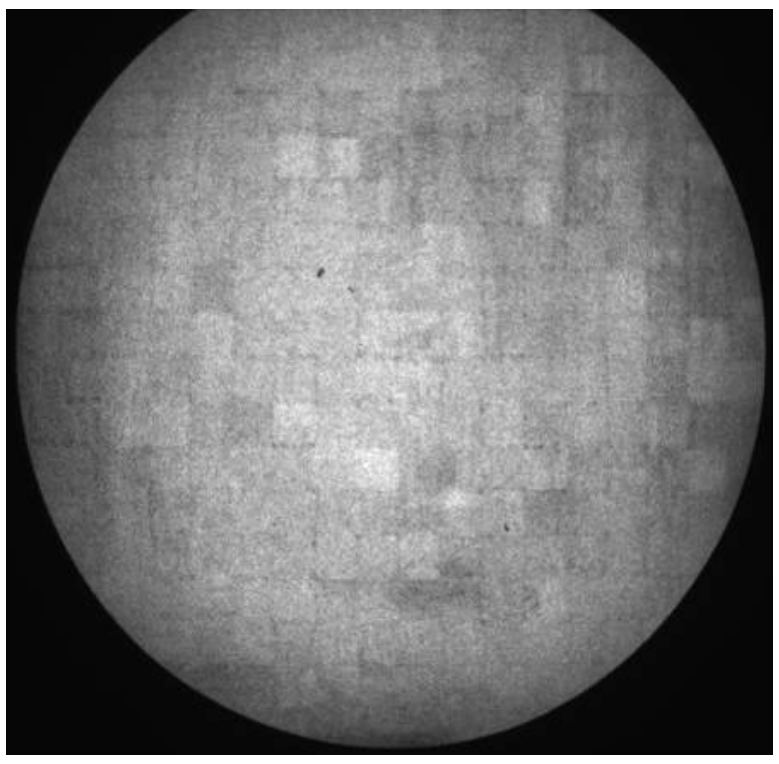

**Fig. 5c Flat image of the fiber screen**

Optik GmbH (PGO) [18]) with the reflective side facing the fiber scintillator. This permits collecting also the scintillation photons emitted in the backward direction, i.e., away from the lens. Fig. 5a shows the scintillating fiber screen while fig. 5b shows a magnified view of one of its sections. As can be seen, the screen is not perfectly constructed, displaying voids and misaligned fiber rows, both of which contribute to the non-uniform light output. The latter is visible in Fig. 5c, which shows a full-transmission neutron radiography image (henceforth denoted “flat” image).

The local net sensitive area was evaluated for different regions of the screen and was found to range between 59 % and 72 %. The mean net sensitive area is 63 %.

The light output of a typical plastic scintillator based on polyvinyl-toluene (BC-400 or EJ-200) is about **10,000 photons** per MeV for a minimum ionizing particle. For a polystyrene based fiber-scintillator (BCF-12), the manufacturer quotes a light yield of **7000-8000 photons** per MeV for a minimum ionizing particle [17]. The lower emission value compares unfavorably with inorganic scintillators such as NaI(Tl) or CsI(Tl), which emit 38000-54000 photons per MeV [17] but have much longer decay times.

The emission of light from the fiber was simulated using the ZEMAX [19] optical design program. The simulation indicates that the light is emitted into a shape shown in Fig. 6 and the light trapping efficiency along the fiber is 4.5 % in each direction.

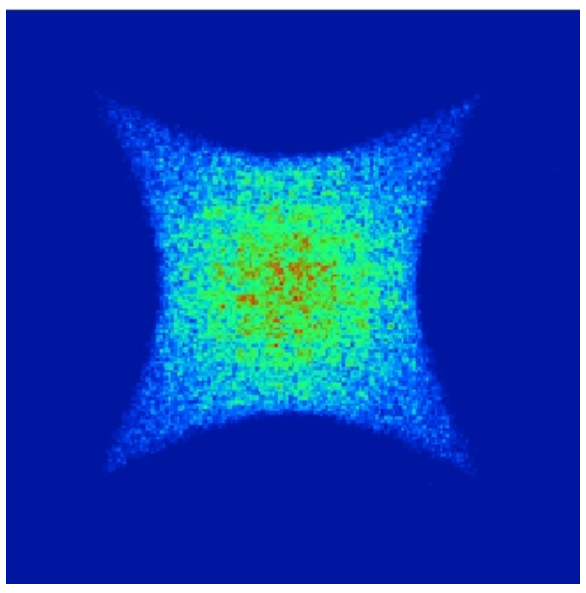

**Fig. 6 Shape of the light beam emitted from the fiber. Increased brightness indicates higher intensity**

### *2.1.1.2 Bending mirror*

In order to protect the electro-optical components (CCD and image-intensifier) from radiation damage, they must be positioned out of the neutron-beam. Thus, a bending mirror is positioned at 45° facing the scintillator screen in order to deflect the scintillation light toward the F#=0.95 lens.

This mirror (3.3 mm thick), also manufactured by PGO [18], is composed of a borosilicate substrate, front-coated by a multi-layer dielectric thin-film. It exhibits a reflectance of >99% at wavelengths up to 700 nm (see Fig. 7), mechanical robustness and high temperature stability.

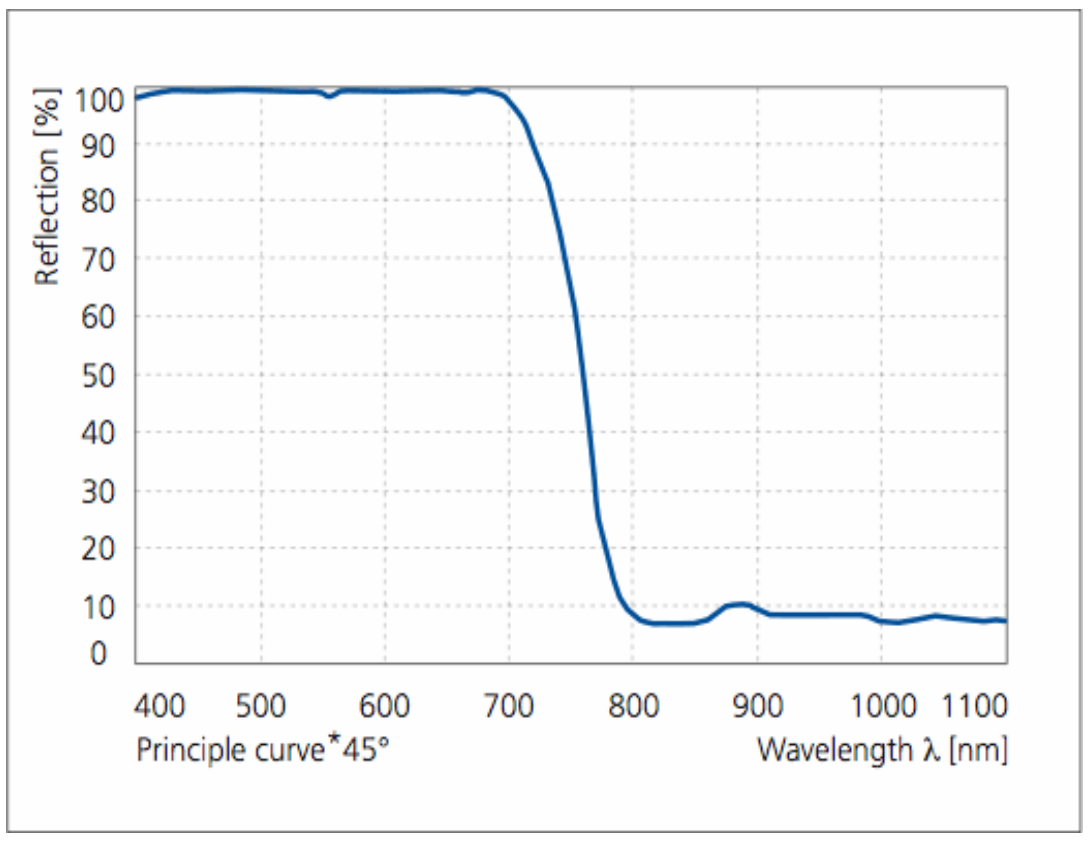

**Fig. 7 Reflectivity curve as function of wavelength [18]**

### *2.1.1.3 Large-aperture lens*

The amount of scintillation-light photons created in a single fiber of the screen per detected 7.5 MeV neutron is rather small, i.e., 5424 photons on average. It is thus important to collect this light on an image intensifier as efficiently as possible. This is done with the aid of a large-aperture lens, 120 mm focal length with a relative aperture F# = 0.95.

Since such a lens was not commercially available, it had to be custom designed and manufactured for this application. A schematic view of this lens is shown in Fig. 8.

The lens consists of 5 elements, 3 positive and 2 negative. Its effective focal length is 120 mm, the relative aperture F# is 0.95 and its entrance pupil diameter is 126 mm.

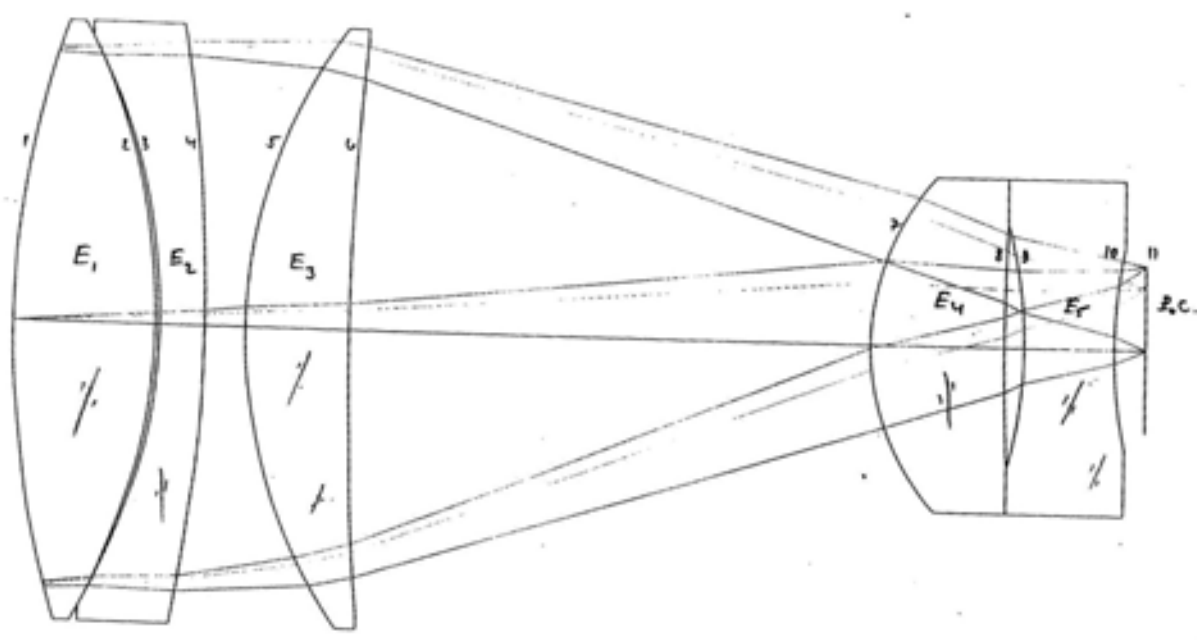

**Fig. 8 Schematic drawing of the collecting lens, 120 mm focal length, F#=0.95**

The lens elements were manufactured by A. Optical Components Ltd., Azur, Israel, and the lens was assembled at Soreq NRC. Figs. 9 & 10 show engineering illustrations of the lens with the image-intensifier coupled at its rear-end (right hand side in the Figs.).

The fraction of light collected by the lens (positioned at distance of 750 mm and exhibiting ~90% light transmission) out of total light emitted from the fiber was also determined by the ZEMAX simulation mentioned in section 2.1.1.1 and found to be **1.3 %.**

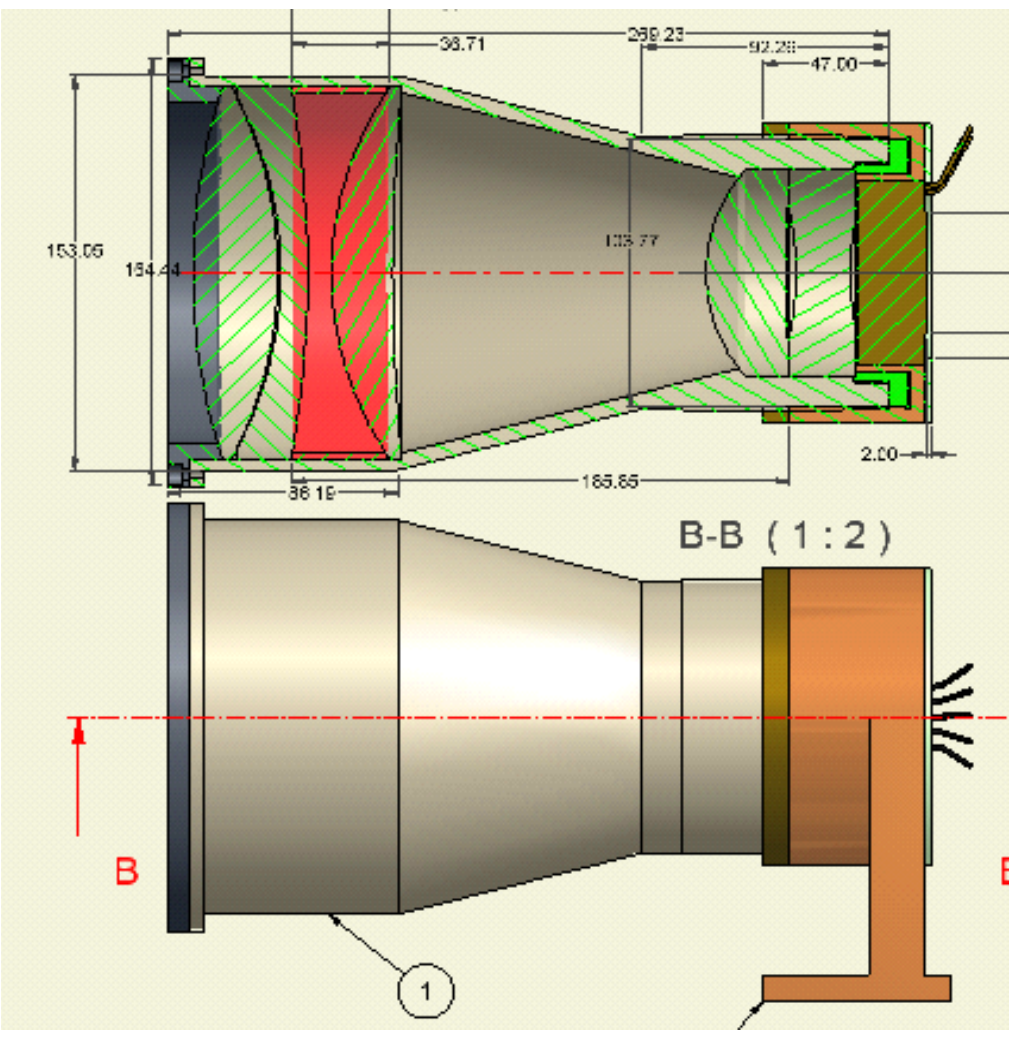

**Fig. 9 Side view of the F#=0.95 lens. Top: cross-section view of the lens and image-intensifier coupling. Bottom: outside view of the lens and intensifier holder (brown)**

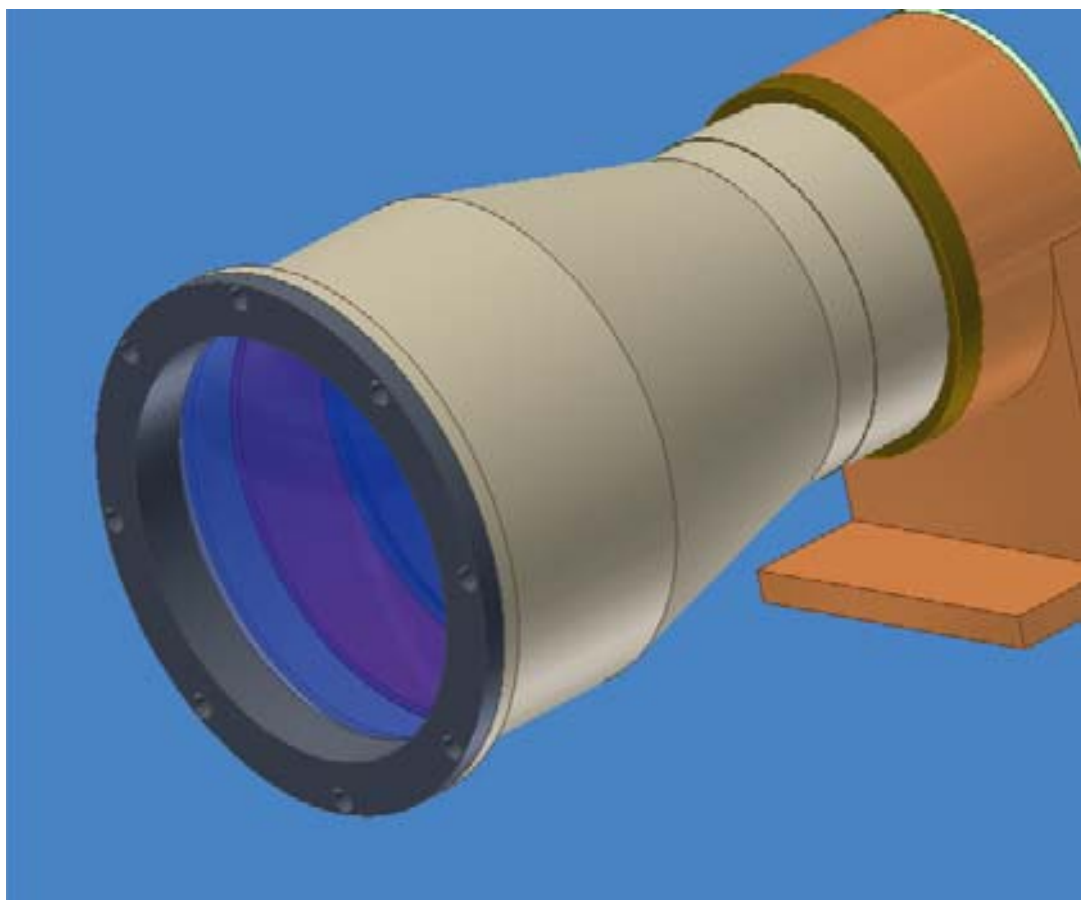

**Fig. 10 General view of the lens**

### *2.1.1.4 Image intensifier*

Energy selection via the time-of-flight technique is achieved by gating the image-intensifier photocathode with high voltage pulses on a ns-duration time-scale. The image intensifier acts as a very fast optical shutter, synchronized with the accelerator pulses and triggered to open after a pre-selected neutron time-of-flight that corresponds to a specific neutron energy.

The image-intensifier (I-I) employed here, shown in Fig. 11, is a high-gain, proximity-focus device. The tube has a circular sensitive area of 40 mm in diameter with a rugged metal ceramic construction. The photocathode is a low noise S20 (maximum sensitivity at 450 nm) with a conductive mesh undercoating to improve pulsing capability. Its quantum efficiency at a wavelength of 420 nm is 13 %**.** Two micro-channel plates enable a gain of $10^6$ W/W. The input window is made of fused silica and the output window is made of a fiber-optic face-plate, plated with a P43 phosphor screen. Its limiting resolution is 20 lp/mm.

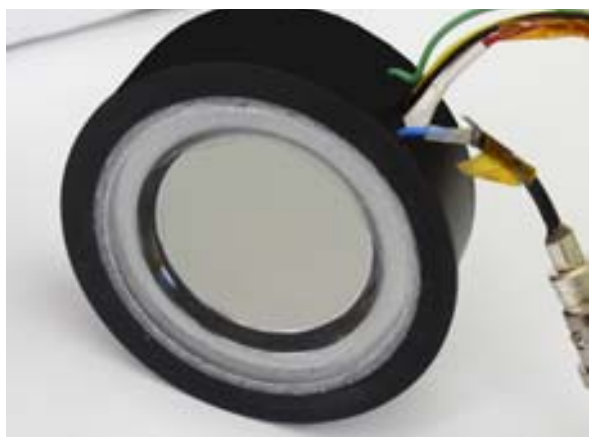

**Fig. 11 The image-intensifier front view**

The spectral response of the image-intensifier S-20 photocathode is displayed in Fig. 12. The maximal sensitivity is 47.26 mA/W at 450 nm, corresponding to a quantum efficiency (QE) of about 13 %.

One of the reasons for the relatively low QE of the photocathode is the method of cathode preparation, which is aimed at reducing the irising effect (that occurs due to the low electric conductivity of typical multi-alkali photocathode materials) within nanosecond gating times. Voltage pulses applied to the rim of the photocathode require significant time to propagate to its center. Fast (ns) gating requires a low resistivity contact which distributes the signal quickly to the entire surface of the photocathode. Such contacts are realized either by a thin metal layer underneath the photocathode or, for even faster gating times (~1 ns), a thin wire mesh.

The tube used in this configuration is of mesh type, gateable down to 3 ns. This mesh absorbs ca. 20 % of the incoming photons before they strike the photocathode. The minimal gate width we achieved for full intensifier opening was determined experimentally to be about 10 ns [20].

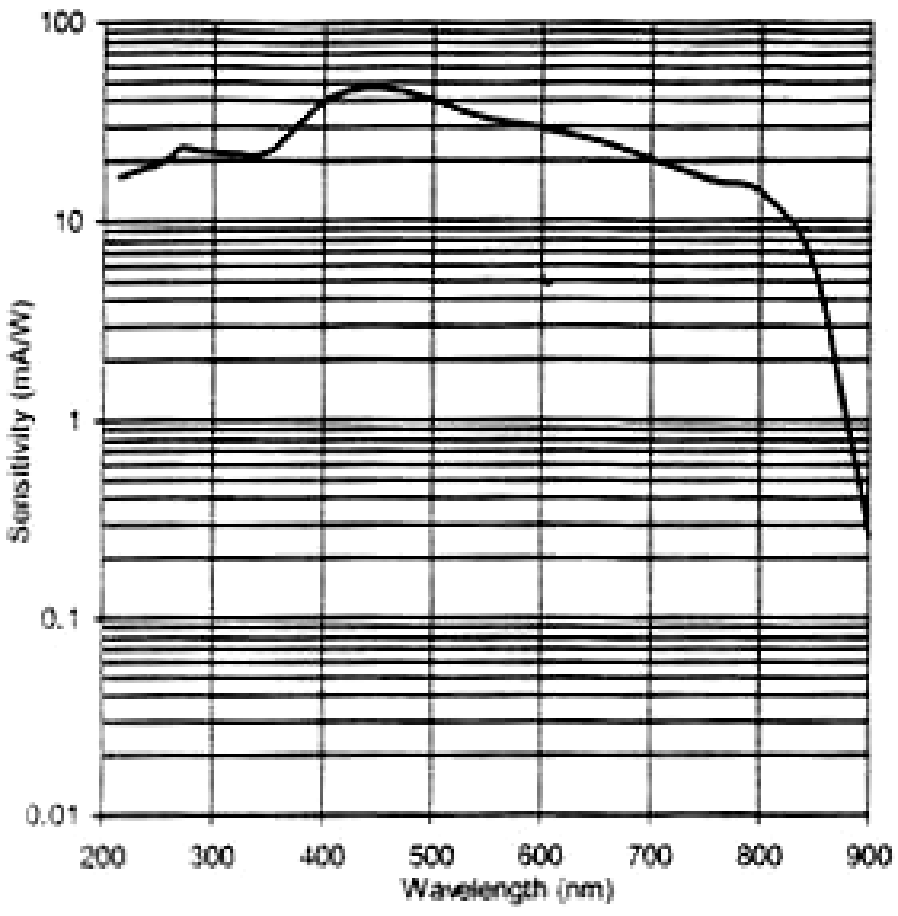

**Fig. 12 Spectral response of the LN S20 photocathode**

In order to perform energy spectroscopy using the TOF technique, the I-I photocathode has to be gated at a negative potential relative to the MCP cathode at a precise delay and width following the accelerator burst. For this purpose, a computer-controlled Gate and Delay Generator (G&DG) and a fast High-Voltage (HV) gate pulser were developed and optimized for our application. The G&DG was developed and built at PTB and the HV pulser was custom designed and purchased from Roentdek, Germany [21]. Fig. 13 shows a block diagram of intensifier pulsing (top) and the pulsing regime (bottom) [22].
The G&DG is a single NIM module computer-controlled via a RS232 port. The gate and delay widths can be obtained in coarse time steps of 8 ns, derived from a 125 MHz clock. The fine tuning is realized by digital delay lines providing steps of 0.25 ns. To avoid time jitter between the G&DG clock and the accelerator beam pulse, the 125 MHz clock was phase-locked to the cyclotron frequency.

The output pulse from the G&DG triggers the Roentdek HV-gate pulser [23]. The output switches the photocathode voltage between +50 V (intensifier-off) and –150 V (intensifier-on) with rise-times of 2-5 ns (depending on the capacitive load). The high switching frequency (2 MHz) and the large capacitance of the Field-Effect-Transistor (FET) and FET pulser-driver cause significant heating due to high power deposition (about 30 W dissipated power). This heat is removed by coupling the pulser box to a water-cooled copper block (see Fig. 14). Minimum achieved gate width was 10 ns at the full repetition rate of 2 MHz.

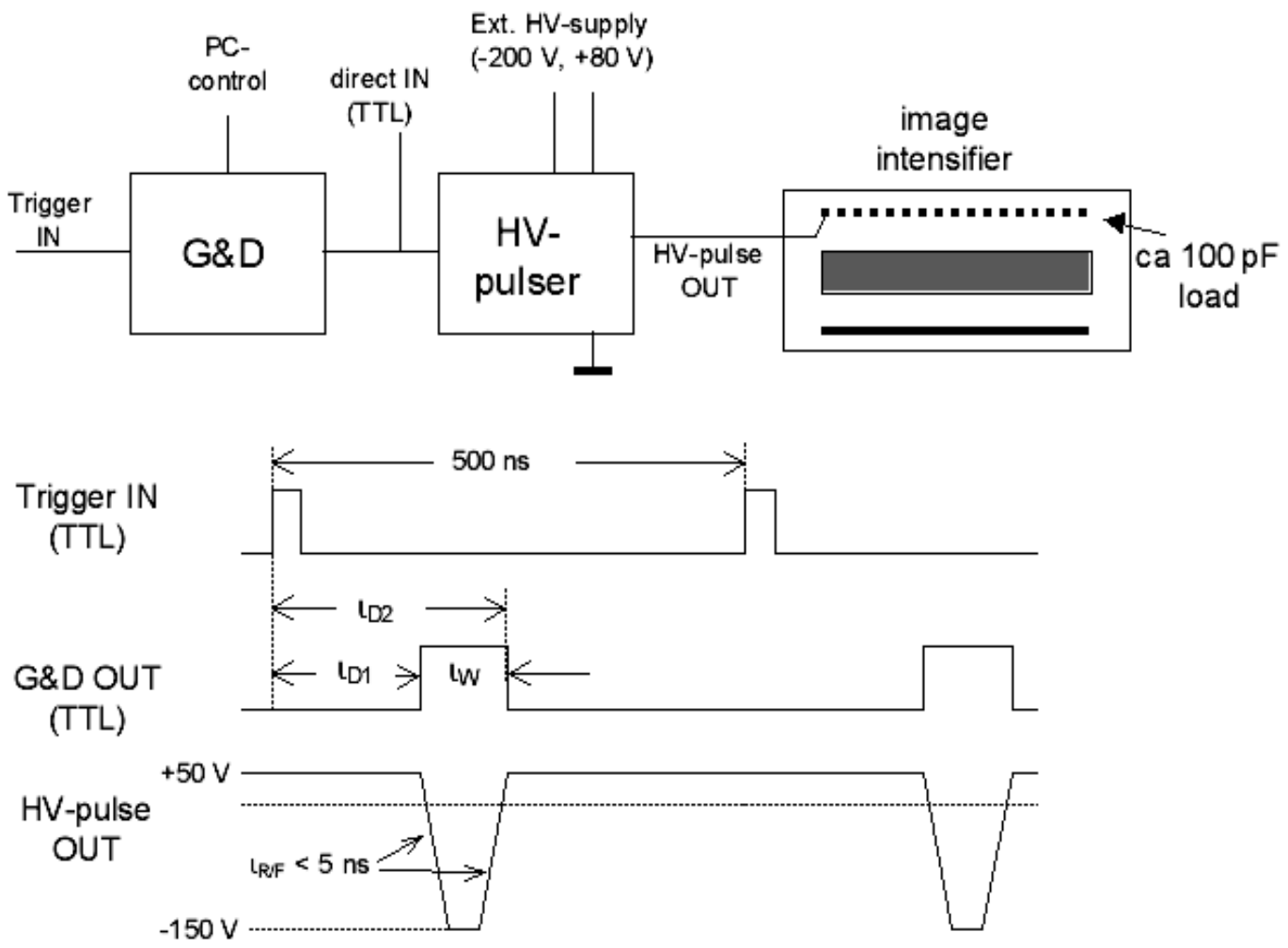

**Fig. 13 Block diagram of intensifier pulsing (top) and pulsing regime (bottom). Delays: $L_{D1}$, $L_{D2}$, gate width: $L_w$= $L'_{D2}$ -$L_{D1}$**

### *2.1.1.5 Tandem lenses*

A tandem configuration of a 200 mm F# = 2.8 Canon lens and a large-aperture Nikon 50 mm F#2.8 is used to relay the image of the intensifier phosphor to the CCD. As can be seen in Fig. 14, the lenses are mounted face-to-face, the object (intensifier) and the image (CCD) being at the foci of the 200 and the 50 mm lenses, respectively. This combination provides 4:1 demagnification (an object 40 mm in diameter is viewed by a 10 mm CCD).
The tandem configuration was chosen because it provides a factor of ~4 higher light detection efficiency than a single lens configuration [24]. A fiber-optic coupling would be more efficient but will not allow easy replacement of the camera or image-intensifier. The disadvantage of the tandem configuration is the introduction of significant vignetting.

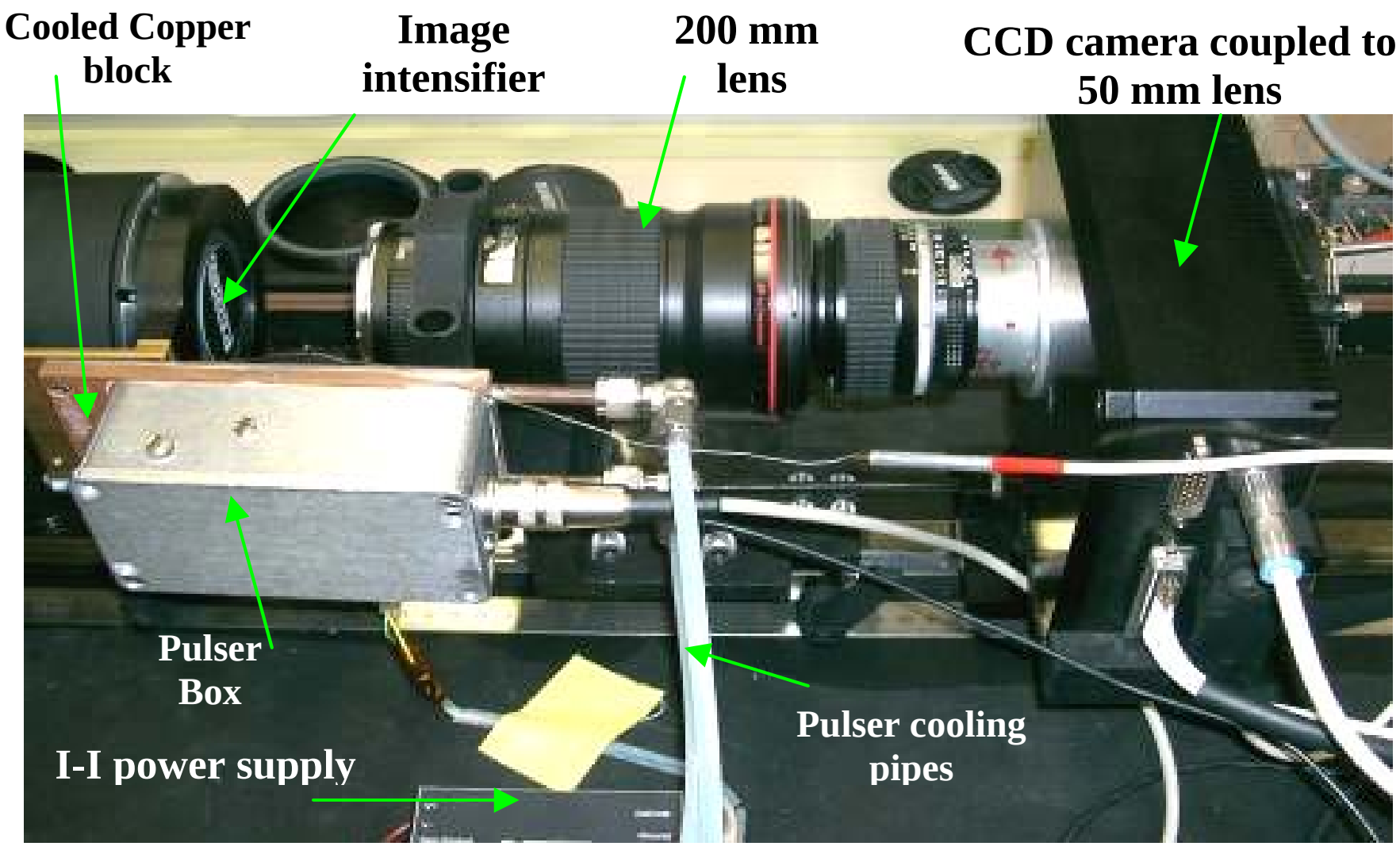

**Fig. 14 Image of the pulser (left) mounted next to the image intensifier and the tandem lens configuration**

#### *2.1.1.6 CCD camera*

The CCD camera used in TRION Gen.1 is the Chroma C3, manufactured by DTA Scientific Instruments, Pisa, Italy [25]. The CCD sensor of the camera was KAF 1602E.

### 2.1.2 Imaging results

Figs. 15 to 16 show radiography images obtained with TRION Gen. 1. A more comprehensive analysis of TRION's performance can be found in refs. [20, 26]

#### *2.1.2.1 Imaging with broad-energy neutron spectrum*

The performance of the TRION detector was tested with a ns-pulsed (beam-burst ~1.5 ns wide, pre-selectable repetition rate in MHz range), variable-energy neutron beam produced by the PTB cyclotron. The cyclotron is equipped with a fast internal beam-pulsing system, that provides the capability to choose between a wide range of pulse repetition rates, 2 kHz to 6 MHz [27]. The actual rate had to be restricted to 2 MHz (500 ns separation between consecutive pulses) in order to prevent the most energetic neutrons produced in a particular beam burst from arriving at comparable times to the less-energetic neutrons produced in the preceding pulse. The reaction employed to create the fast neutrons was the $^{9}$Be(d,n) ($E_d$ = 9 MeV). The average deuteron beam current at 2 MHz was 1.5-2 μA with spot size of 3 mm. The emitted neutron energy can range between 0 – 14 MeV. A more detailed description can be found in refs.[20, 26].

Figs. 15 a & b show the fast-neutron radiograph and a photographic image of a phantom, respectively. The TRION detector was positioned 12 m away from the neutron source while the phantoms were ca. 30 cm from the detector. For the neutron image, the full spectral distribution of the neutron beam was utilized, excluding the gamma peak by setting an appropriate gate in the TOF spectrum.

The phantom consists of a plastic toy gun, a trumpet mouthpiece, two vials containing water and acetone mixtures (numerical markings on the vials indicate water volume percentage) and two regular AA batteries wrapped with electrical wire. As can be seen, image resolution is good enough to allow visual inspection, as with X- and γ-ray images. But, unlike the latter, which would only exhibit strong contrast for high-Z materials (such as the brass mouth-piece), the fast neutron image displays high contrast for low-Z materials as well. Fig. 15c shows the Contrast-Transfer Function (CTF) curve for TRION Gen.1, displaying quantitatively the spatial resolution for different spatial frequencies.

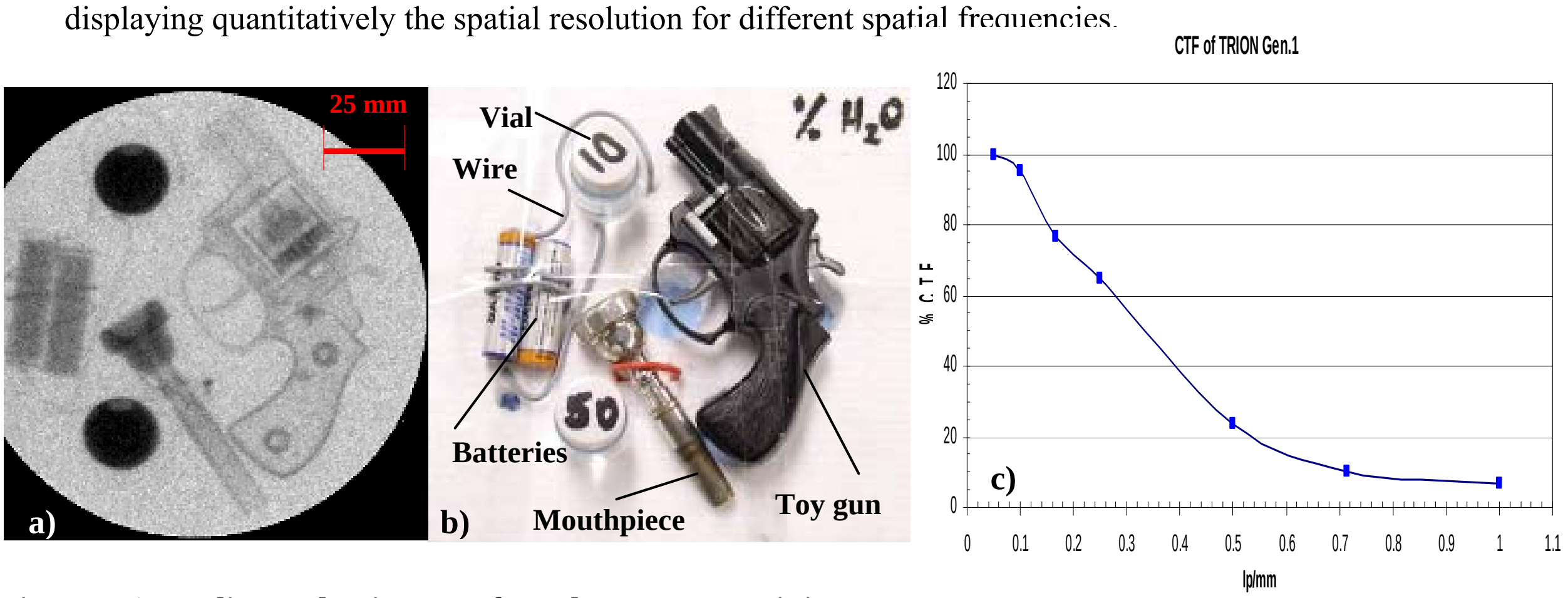

**Fig. 15 a) Radiography image of a phantom containing a plastic toy gun, a trumpet mouthpiece, vials containing water and acetone mixtures, batteries wrapped with electrical wire; b) A photographic image of the actual phantom; c) CTF curve of TRION Gen. 1**

Figs. 16 a – c display a phantom consisting of a plastic toy gun, a hollow tungsten block, two vials containing water and acetone mixtures (partial water volume in the vials is 50% and 10%), melamine powder and a BNC connector. To study image quality when the phantom is shielded by a dense high-Z material, this phantom (Fig. 16b) was placed behind 1" thick lead bricks (Fig. 16c) and radiographed (Fig. 16a). Vial length was 55 mm and melamine sample thickness was 50 mm. As can be seen, the spatial resolution appears to be unaffected by the presence of the lead bricks. The good contrast visibility of the low-Z materials is also evident. These images show that TRION can produce high resolution fast-neutron transmission images that can provide, like gamma images, the means for visual identification of suspicious items within inspected baggage. Moreover, low-Z materials appear with high contrast and cannot be easily masked by high-Z shielding material.

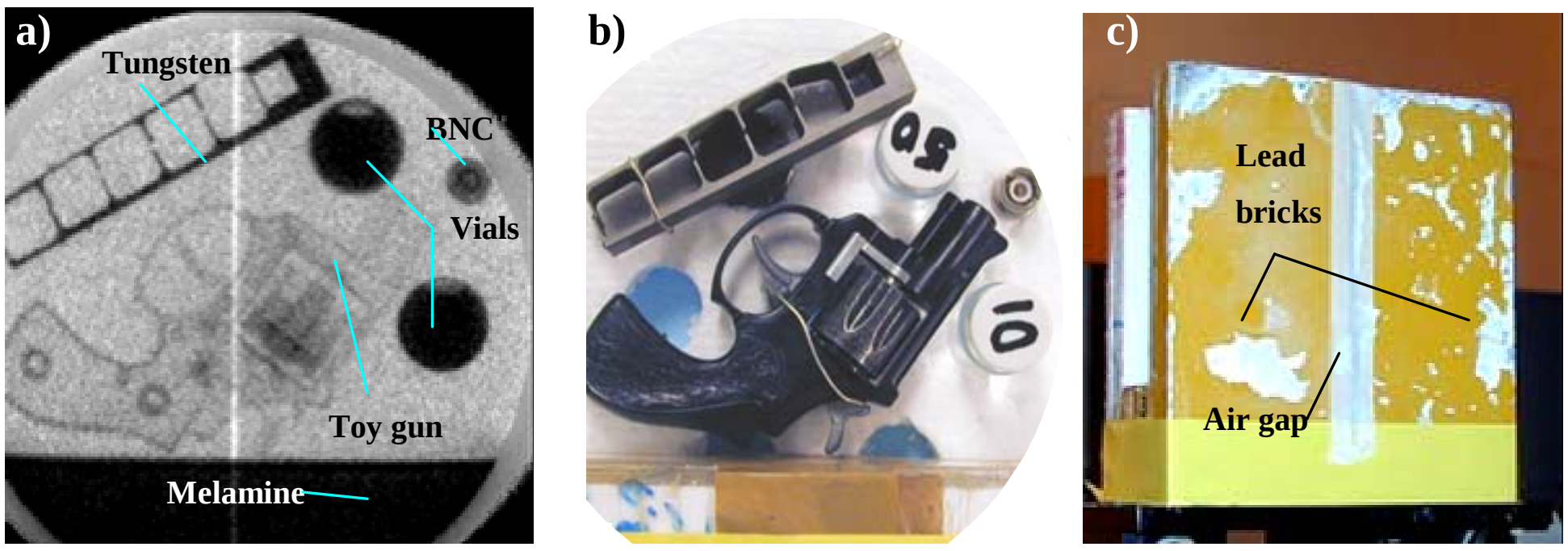

**Fig. 16 a) Radiography image of a phantom; A photographic image of the actual phantom (b) placed behind 1-inch-thick lead bricks (c). The gap between the bricks shows up in (a) as a white vertical line**

### *2.1.2.2 Elemental imaging*

Fig. 17 shows the fast-neutron flux attenuation vs. neutron TOF for a 10 cm block of graphite and 22 cm thick liquid nitrogen absorber. The characteristic spectral features of each element can be used for identifying and determining its elemental content.

Fig. 18 shows an object consisting of melamine (which simulates here a commercial nitrogen-rich explosive), several carbon rods and a steel wrench, that was consecutively imaged at various neutron energies. The top row of Fig. 18 shows 6 neutron images, marked with the corresponding TOF windows. With no previous information, they all look quite identical; however, with appropriate least-squares fit solutions, the net C and N distribution of the objects can be derived (Fig. 18, lower right). Also, as expected, the steel wrench disappears completely in this process.

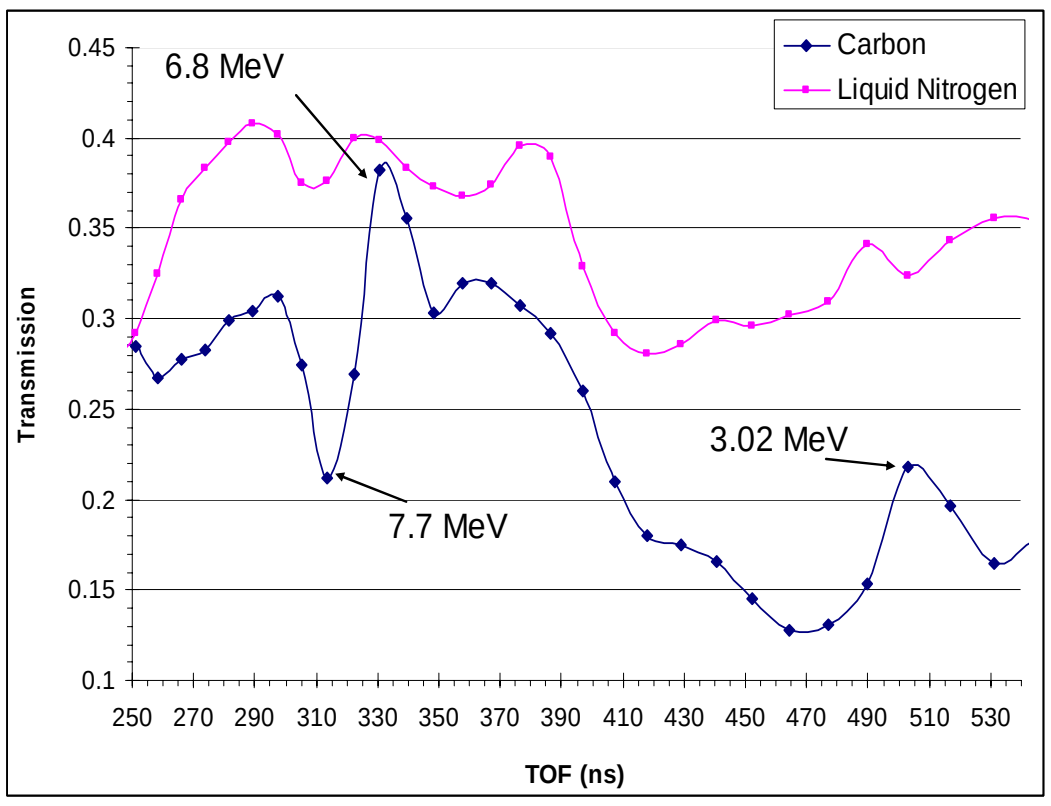

**Fig. 17 TOF spectra of d-Be fast neutrons transmitted through a block of graphite and liquid nitrogen, obtained with TRION Gen.1**

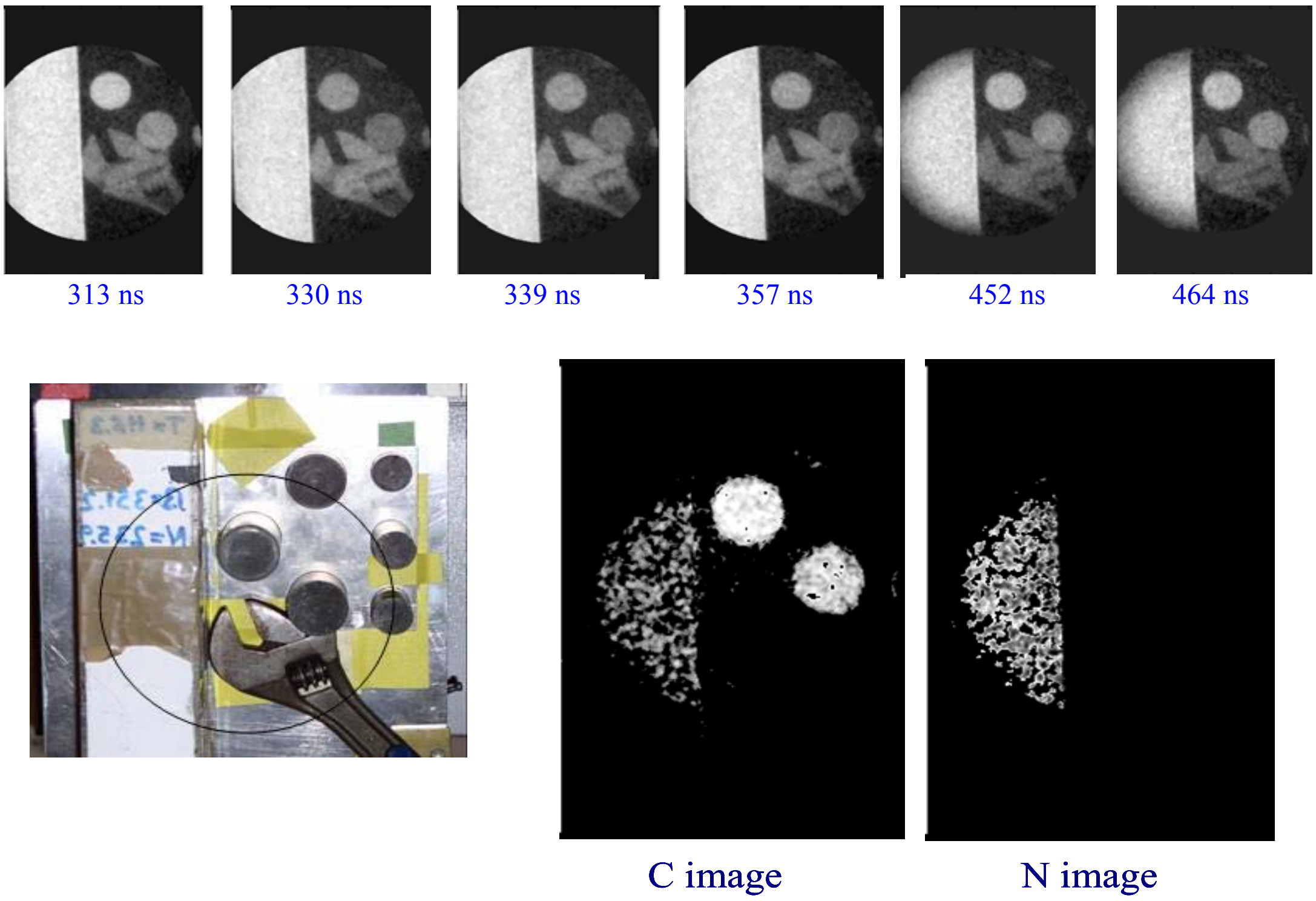

**Fig. 18 Example of resonance imaging using TOF for energy selection. The bottom left image shows a photo of the sample. The circle marks the region which was imaged by the neutron detector (ca. 10 cm diameter). The top row of shows 6 neutron images taken at different energies (indicated is the corresponding TOF below each image). From these images the net carbon and nitrogen distributions in the sample are derived (see the two images at bottom right). Melamine contains ~28 % carbon.**

In order to allow extraction of elemental information from a series of time-resolved radiographs, as seen in fig. 17, the neutron energies should be chosen such that an image should be taken where a significant difference in the cross-section for a single element exists. For nitrogenous material, for example, some of the energies of importance are: 3.6, 4.2, 4.9 and 7.7 MeV.

## 2.2 Concept and results with second generation multi-frame system

### 2.2.1 Components and system optimization

With TRION Gen.1, TOF multi-frame imaging is performed sequentially, i.e., for each TOF frame, the intensifier is triggered to capture just one specific energy interval. This procedure is time consuming and inefficient in its use of the broad-energy neutron spectrum. Thus, in order to progress towards a real-time operational system, it is necessary to acquire images for several TOF frames (or corresponding pre-selected energy regions) simultaneously. This can be achieved by employing several ns-triggered intensified CCD cameras, such that each camera acquires a transmission image corresponding to a different energy region. Fig. 19 shows two views of TRION Gen. 2, containing 4 optical channels.

The front section, comprising the fiber scintillator screen, mirror and F#0.95 lens remained the same as in TRION Gen.1. However, the large-area gated image-intensifier was replaced by an ungated intensifier with a fast phosphor used as an Optical Pre-Amplifier (OPA).

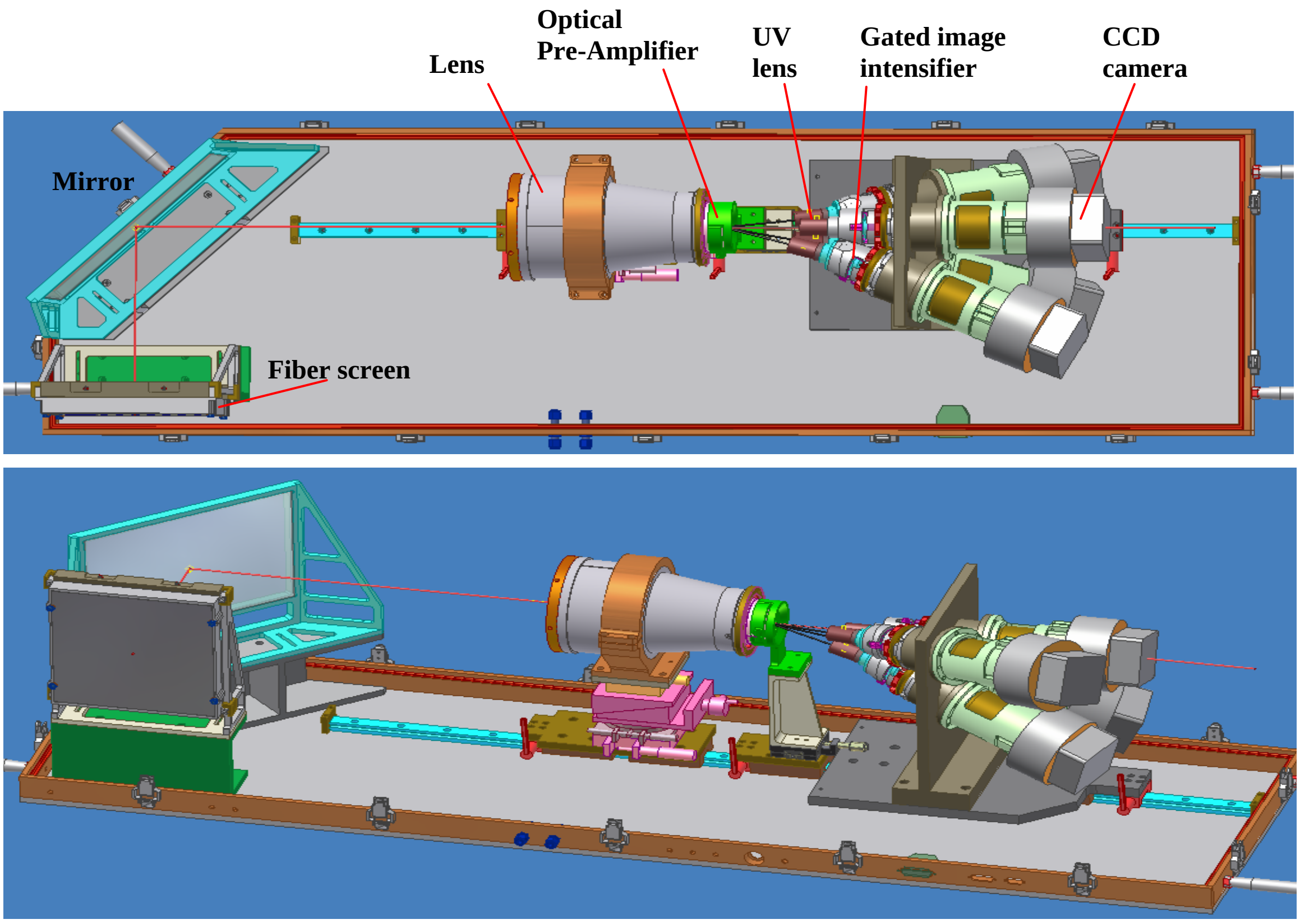

**Fig. 19 TRION Gen. 2 from two views: top (above) and side (below)**

As the light image from the OPA is emitted in the near-UV range, custom-designed UV lenses were employed to relay this light to gated image-intensifiers, incorporated in each of the 4 optical channels.
Fig. 20 shows the rear section of TRION, containing the fully- equipped 4 optical channels.

The assembly contains a central optical channel aligned on the principal optical axis, and 3 other channels that view the phosphor screen of the OPA at various orientations around the optical axis. Each optical channel contains a gated, small (18 mm diameter) image-intensifier (manufactured by Photonis-DEP [28]) that captures the image at the OPA phosphor screen after a preselected neutron TOF and relays it, after further amplification, to a cooled CCD camera (ML0261E, manufactured by FLI [29]).

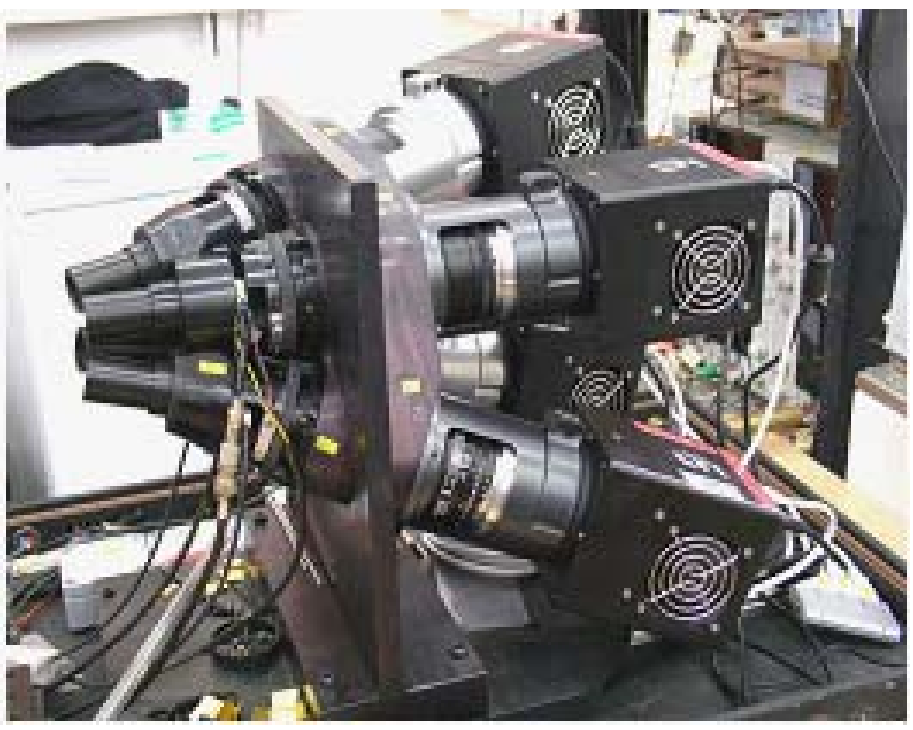

**Fig. 20 Side view of TRION's rear section containing all 4 optical channels**

### *2.2.1.1 Optical Amplifier (OPA)*

To achieve high photo-detection yield a special, large area (40 mm diameter) image intensifier, employed as an Optical PreAmplifier (OPA) was proposed and jointly manufactured by two companies (Photek Ltd, U.K, [31] and ElMul, Israel [32]). It is of crucial importance that the glow curve of the optical preamplifier be as fast as that of the scintillator screen (of the order of 2 ns FWHM) to preserve the good TOF resolution of the imaging system. While all components of a MCP-based, proximity-focus image-intensifier are very fast (sub ns), the sole critical aspect is the decay time of the phosphor screen. All standard, present-day image-intensifier phosphors exhibit glow curves with primary decay times (not including various long after-glow components) ranging from 100-nanoseconds up to tens of milliseconds [31]. However, El-Mul Technologies Ltd. commercially offers a fast phosphor screen referred to as E36 for open multi-channel plate systems used in TOF electron microscopy. Fig. 21 [30] shows the E36 glow curves. The light decay constant of the phosphor screen is 2.4 ns. The disadvantage of the E36 is its relatively low light output.

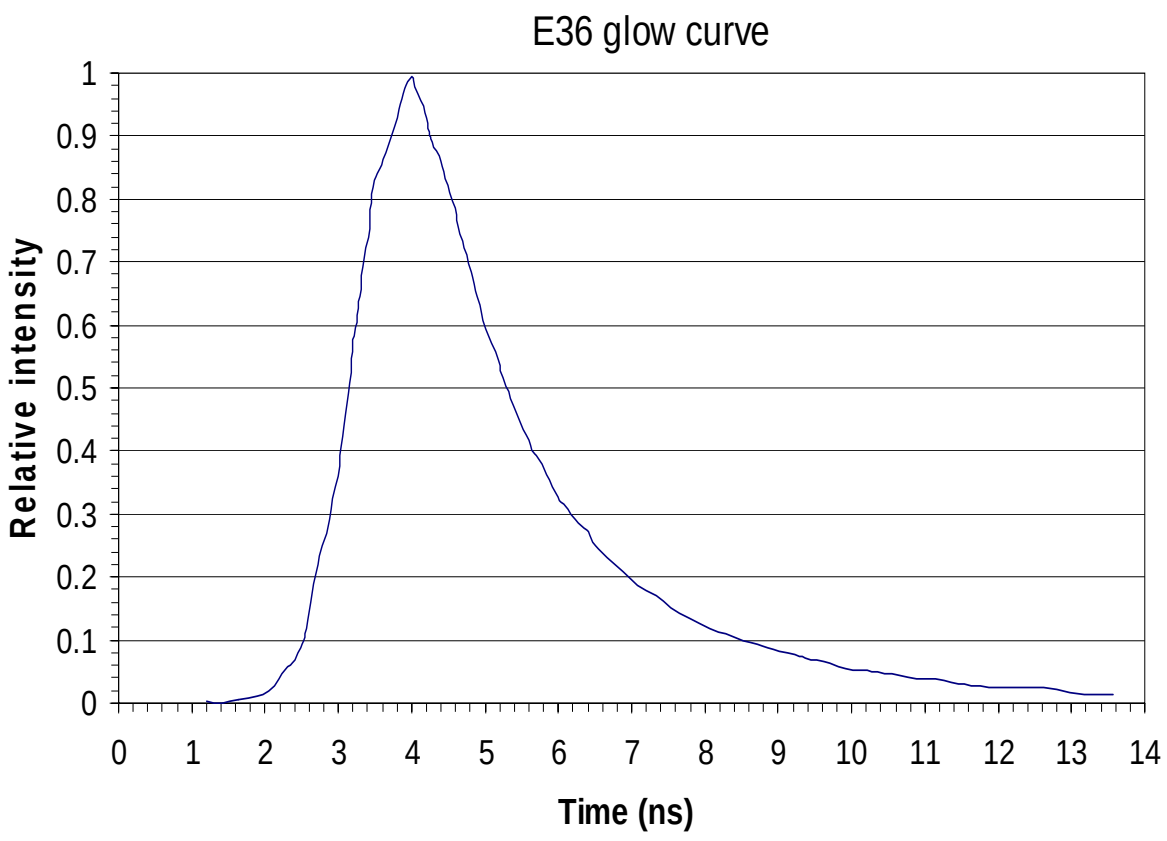

**Fig. 21 Glow curve of the ultra-fast phosphor E36 manufactured by El-Mul [30]**

The specs of the OPA are [33]:

Diameter: 40 mm; photoathode type- Bialkali; QE @ 420 nm is 18 %; quartz entrance window; phosphor screen type (ElMul) E36; electron gain- $5\times10^4$ e/e; phosphor emission range is 370 – 410 nm with maximum at 394 nm; borosilicate output window.

### 2.2.1.2 Improvement of signal to noise ratio

Due to the relatively low intensity of the PTB neutron beam, long exposure times were required. Consequently, the images acquired contained a non-negligible fraction of thermal noise from the photocathode. In order to reduce this noise the photocathode window had to be cooled. The cooling was achieved by blowing dry cold air on the photocathode window. Fig. 22a depicts a schematic drawing of the plastic cooling ring containing 10 air nozzles, while Fig. 22b shows the OPA and the 120 mm lens with the cooling ring mounted on its rear end. During operation, the OPA is in close proximity to the ring. Cooling tests showed that the thermal noise was reduced by a factor of 24 to insignificant levels, yielding a signal to noise ratio of 200.

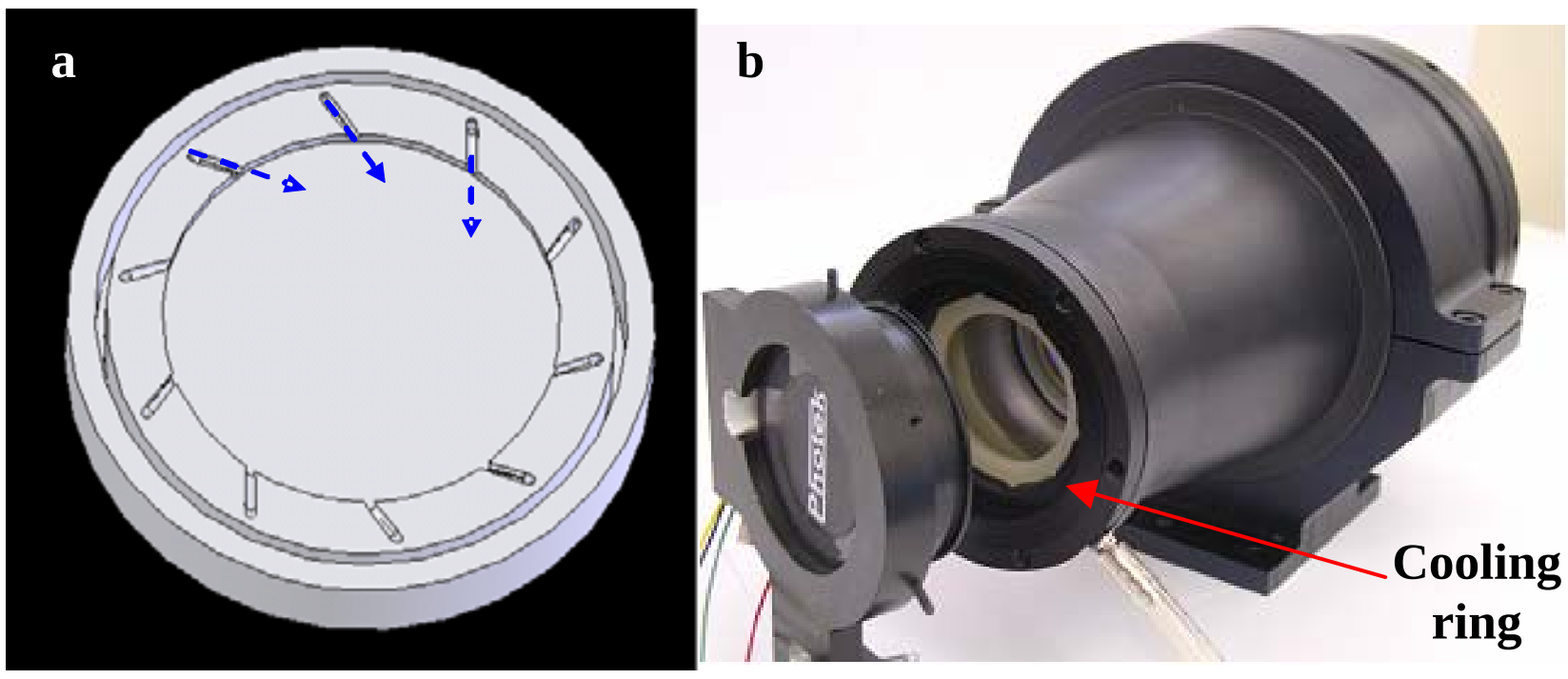

**Fig. 22 a) A schematic illustration of the cooling ring with the direction of air flow marked by blue arrows. b) The OPA (seen on left) and the 120 mm lens. The cooling ring is mounted at the rear end of the F#0.95 lens**

### 2.2.1.3 UV relay lens

As the OPA phosphor screen emits in the near-UV range (394 nm), a UV lens for each of the 4 optical channels was custom-designed and manufactured. This lens is highly efficient (about 85% transmission) at 394 nm; it has focal length of 38 mm and F#3.8. Fig. 23 shows a schematic diagram of such a UV lens, which consists of 3 elements mounted within a special housing.

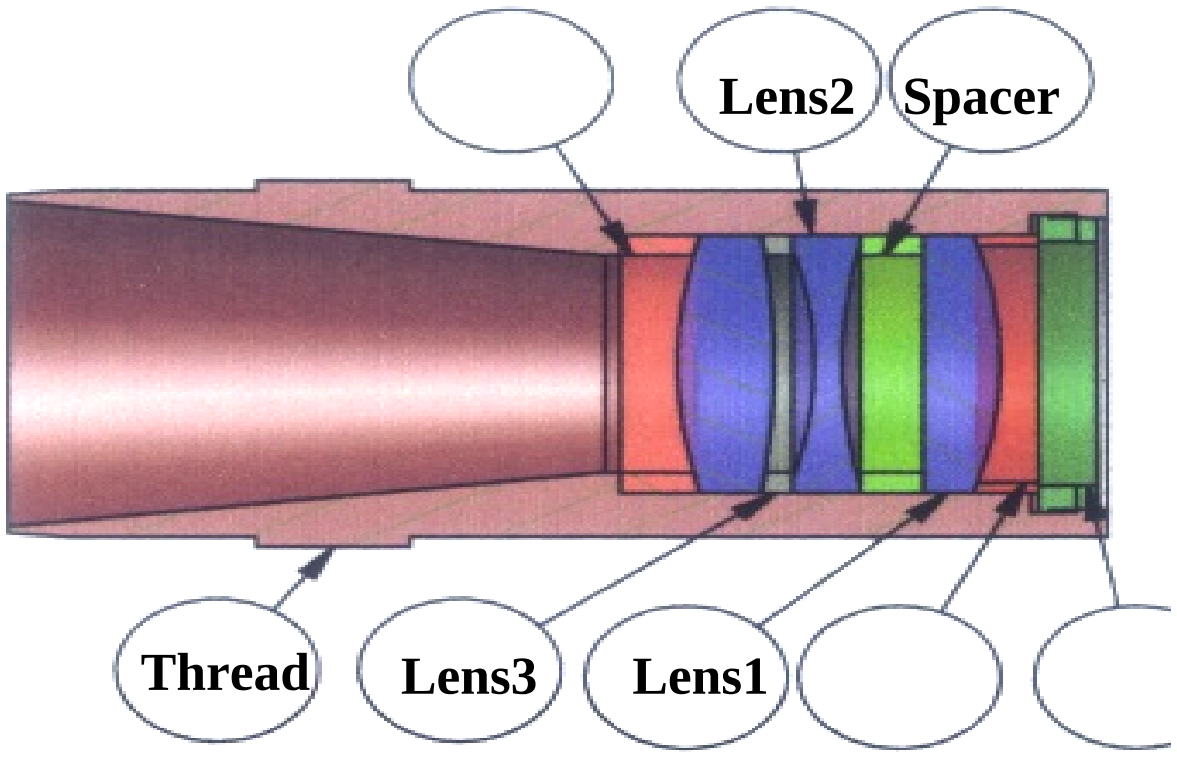

**Fig. 23 Custom-designed UV lens**

### 2.2.1.4 Gated image-intensifiers

For each optical channel, the image from the fast phosphor of the OPA is focused by the UV lens on the photocathode of a gated image-intensifier (I-I). The gated intensifier is an 18 mm diameter, type XX1450AAF produced by Photonis-DEP, Holland [28]. A highly transmissive low-resistance mesh is positioned between the photocathode (S20-UV) and the entrance window of the I-I, to serve as electrical contact, which, in principle, allows gating with pulses as short as 1 ns. With our gating electronics (see section 2.3.1.7, below) and this intensifier we have achieved a gating speed of 4.6 ns at a repetition rate of 2 MHz. The specs of the gated intensifiers are:

Cathode type- S20-UV, Quantum Efficiency (QE) at a wavelength of $\lambda$ = 394 nm is 15 %, phosphor screen type: P43 with fiber optic output window and the photon gain at $\lambda$ = 400 nm is $4\times10^{3}$ emitted photons per incoming photon. The tube is equipped with a single MCP.

### *2.2.1.5 Defocusing correction*

Each of the optical channels views the OPA output screen from a different orientation relative to the principal optical axis. Only the central channel is positioned on this axis, while the optical axes of the other channels are positioned at an angle of 17.5° relative to it.

In the central optical channel, oriented along the principal optical axis, the image planes of the OPA phosphor, the UV lens and the photocathode of the 18 mm I-I are all parallel. In contrast, for the other channels, the image plane of the OPA and those of the UV lenses are tilted at 17.5° to each other. In such a configuration, certain parts of the viewed object (the phosphor of the OPA) will be outside the focal plane of the UV lens, resulting in a partially-defocused and distorted image. This problem can be alleviated (to a large extent) by tilting the image plane (photocathode of gated I-I) using the Scheimpflug principle [34-36]. This principle states that the image plane is rendered sharp when the three planes (object, lens and image) intersect along one line. Hence, if the angle between the object plane and the lens plane is $\alpha$, the image plane should be tilted relatively to the lens plane by an angle $\beta$, such that:

$$\tan(\beta) = m \times \tan(\alpha) \qquad \text{(eq.1)}$$

Where m is lens magnification. In our case $\alpha$ is 17.5° and m=0.45, so the image intensifier should be positioned at 7.8° relative to the lens plane, as seen in Fig. 24.

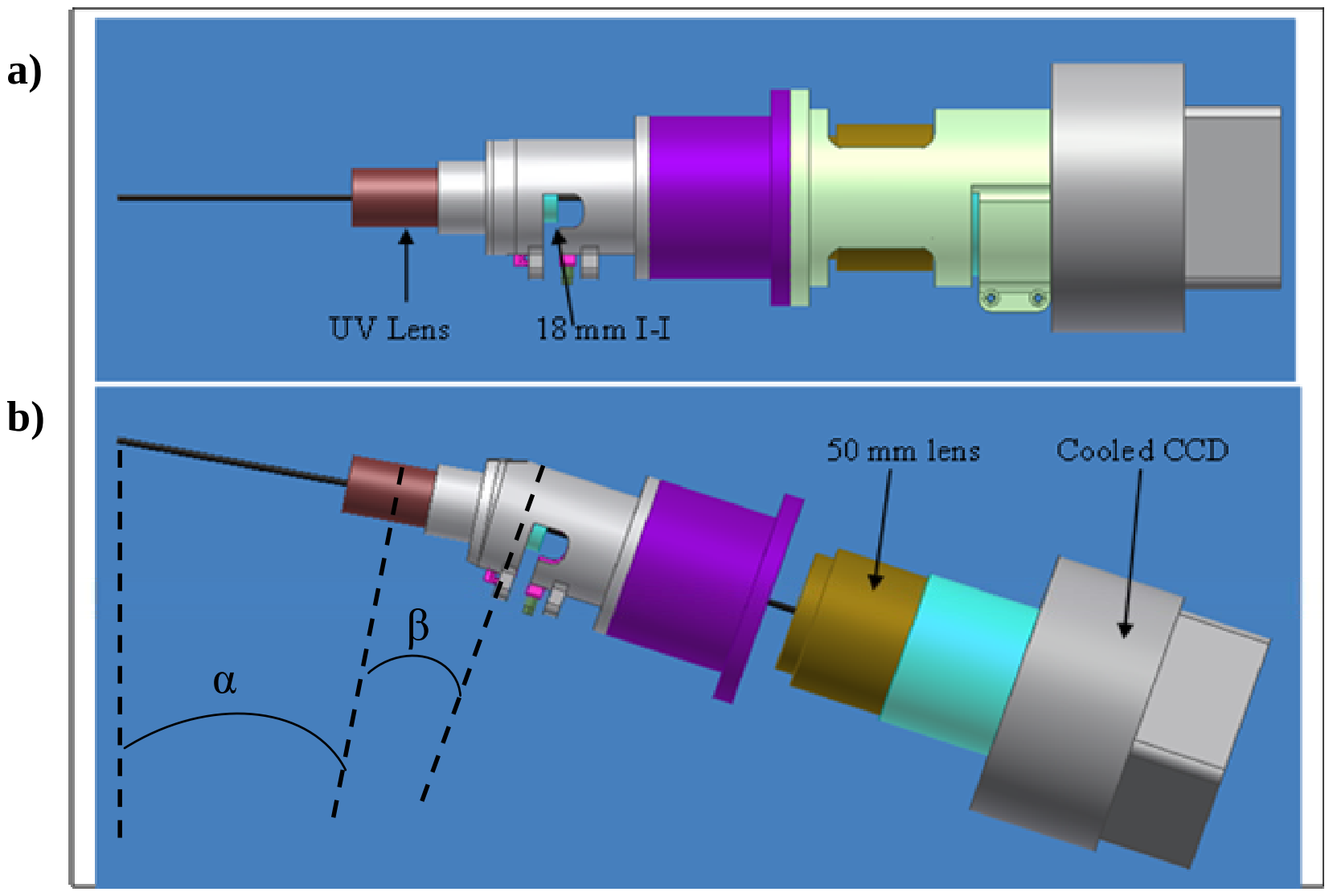

**Fig. 24 a) Central and b) off-axis optical channel**

*2.2.1.6 Off-axis images correction*

Fig. 25 shows images of a grid mask taken by the different optical channels. Fig. 25(0) shows an image taken by the central channel located on the principal optical axis and Fig. 25(1-3) show the different off-axis views. The fact that the off-axis optical channels view the OPA screen at an angle of 17.5° relative to the principal optical axis results in a distorted image. With an alignment program that incorporates MATLAB [37] built-in functions, the off-axis images 1-3 were aligned with the on-axis image 0. Fig. 25(1a – 3a) show the off-axis images after alignment. Pixel-to-pixel correspondence among images acquired at different energies is a pre-requisite for deriving elemental distributions via pixel-by-pixel arithmetic.

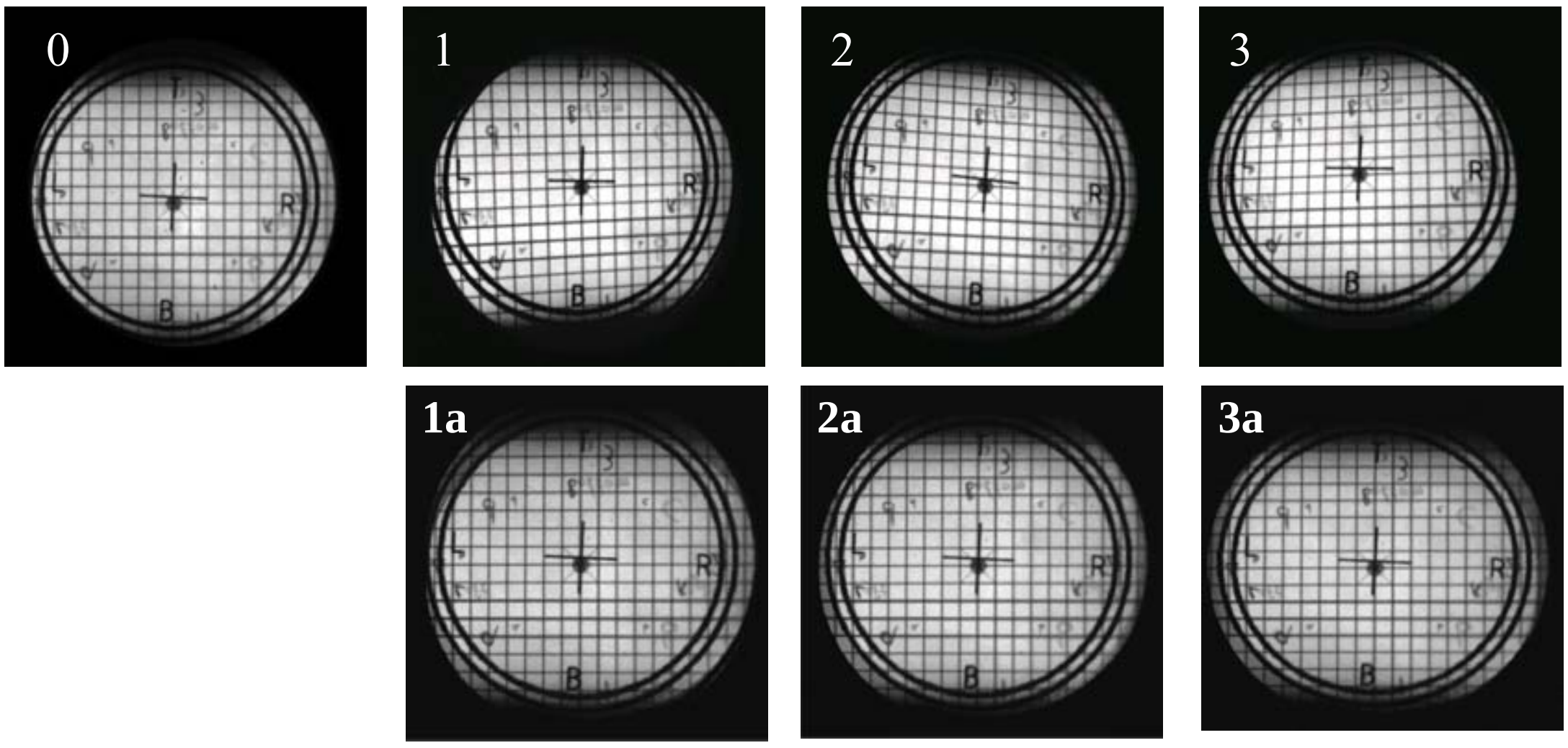

**Fig. 25 0) Central camera on optical axis; 1-3) Off-axis views before correction; 1a-3a) Off-axis images after correction**

*2.2.1.7 Gating of the intensifiers*

The gate pulsers for the four intensifiers were developed at PTB. They are based on an IXYS [38] fast transistor switch and a self-matched clipping-line, as schematically shown on Fig. 26.

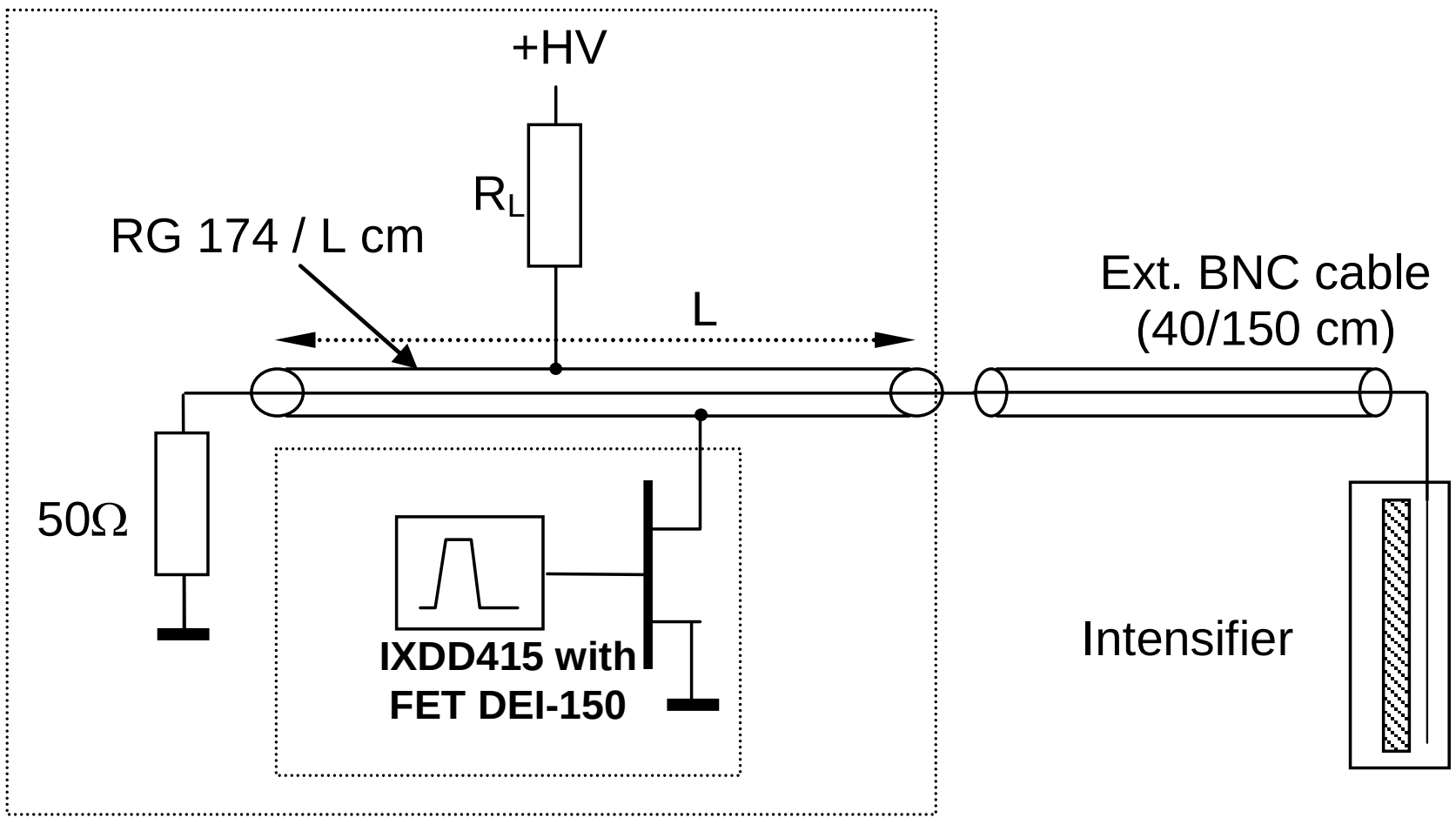

**Fig. 26 IXYS pulser with self-terminating clipping line**

The advantage of this concept is its proper termination at the end of the clipping line. It permits positioning the pulsers at some distance from the image-intensifier. This is important for densely-packed multiple-intensifier configurations, for which it is preferable to keep the electronics outside the optical enclosure.

However, the use of this pulser presented several disadvantages compared to the programmable gate unit used for TRION Gen.1:

- Due to the use of the clipping line, the pulse width Δt is not programmable but fixed by cable length L (Δt = L/20 [ns], L in cm)
- At high pulsing frequencies the baseline drifts slightly. This calls for rather small values of $R_L$, which results in power consumption of about 50 W per gate module at the maximum gating speed of 2 MHz.

During the evaluation of TRION Gen.1, in which ROENTDEK pulsers [21] were used, gate-widths no shorter than 10 ns could be realized. With the new pulsers we have been able to achieve gate-widths as short as 4.2 ns (FWHM) and repetition rates of 2 MHz. The gating of the I-I requires switching the photocathode potential relatively to MCP-IN between +30V to -160V.

#### *2.2.1.8 CCD cameras*

The CCD camera used in TRION Gen.2, ML0261E, was manufactured by Finger Lake Instrumentation (FLI) [29], USA. The CCD sensor is a KAF 0261E. All four cameras were controlled simultaneously via the computer using a custom made program with integrated FLI drivers.

### 2.2.2 Imaging results

For the purpose of demonstrating the quality of gamma-ray and neutron images using the TRION Gen.2 detector, a phantom consisting of the following items was radiographed at the PTB facility: 7.65 mm Walther PPK gun, magazine with gas filled bullets, hollow tungsten bar, UO2 powder.

Each of the 4 cameras captured an image for a time window gate-width at a different delay time (energy) corresponding to the following:

- Camera 0 – Gamma-ray spectrum from the $^{9}$Be(d,n) reaction [39]
- Camera 1 - **10.5 MeV** neutrons
- Camera 2 - **7.3 MeV** neutrons
- Camera 3 - **3.1 MeV** neutrons

Fig. 27 shows a TOF spectrum resulting from the d-Be reaction, measured at the PTB accelerator facility using TRION Gen.2. The 4 vertical red lines in the figure indicate the locations of the above-mentioned 4 energies on the TOF plot.

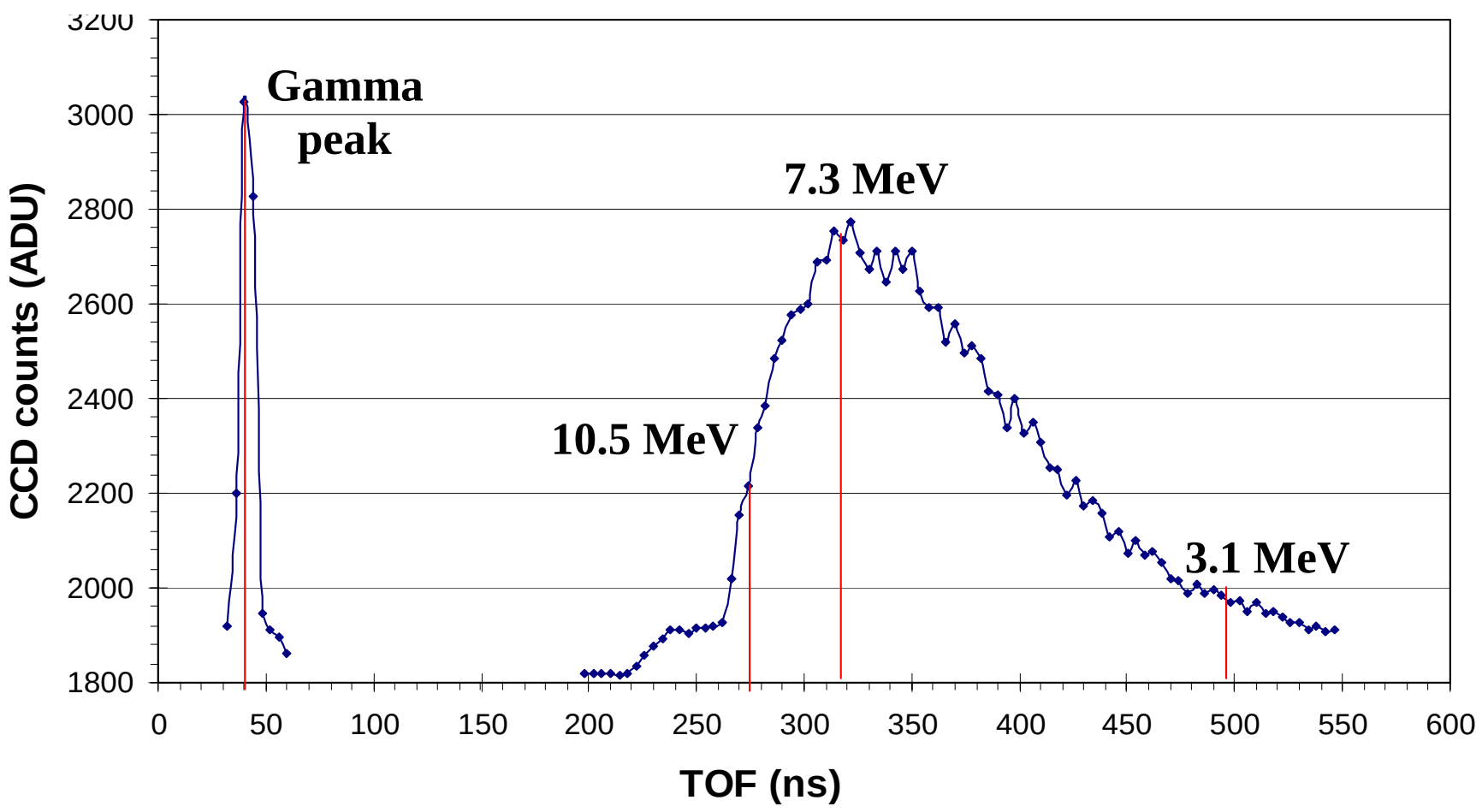

**Fig. 27 TOF Spectra resulting from the d-Be reaction, measured with TRION Gen.2**

Each of the images was normalized by a full-transmission (flat) image taken at the relevant time window. Fig. 28a & b-d show the gamma-ray and neutron images of the phantom. As with TRION Gen.1, the quality of the images is sufficient to permit visual examination of the radiographed objects by an operator. It can be seen that Figs. 28b & d suffer from excessive quantum noise due to the low neutron statistics at those energies [27] (see also Fig. 27). In addition, 3 MeV neutrons create less scintillation photons per detected event than higher energy neutrons.

These results show that the 4 optical channels can be individually operated, each capturing a different time window in the TOF spectrum with very good spatial and temporal resolution.

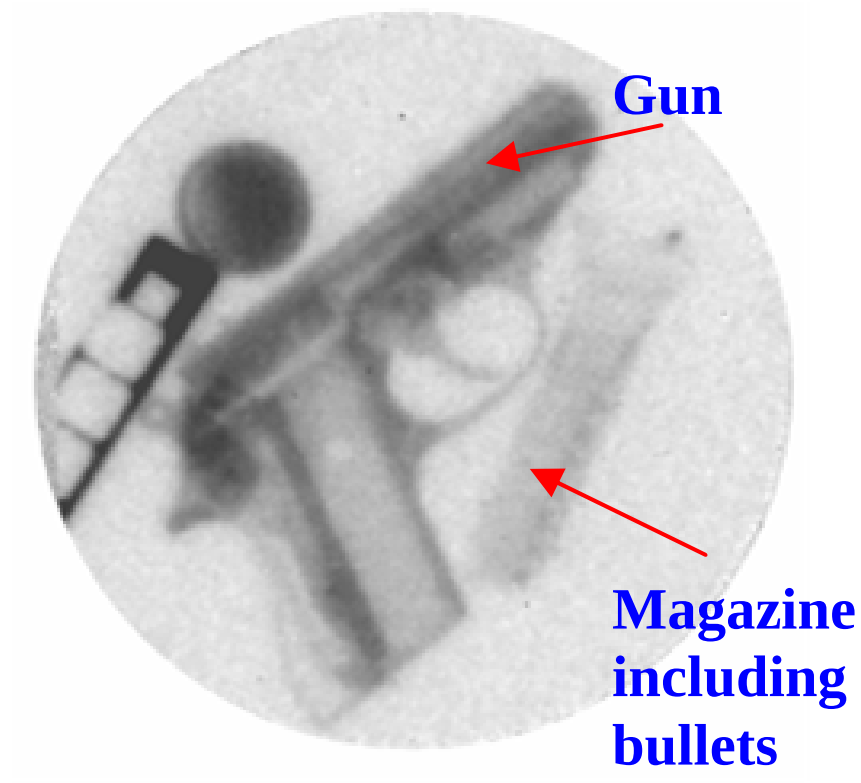

**Fig. 28a Camera 0, gamma image**

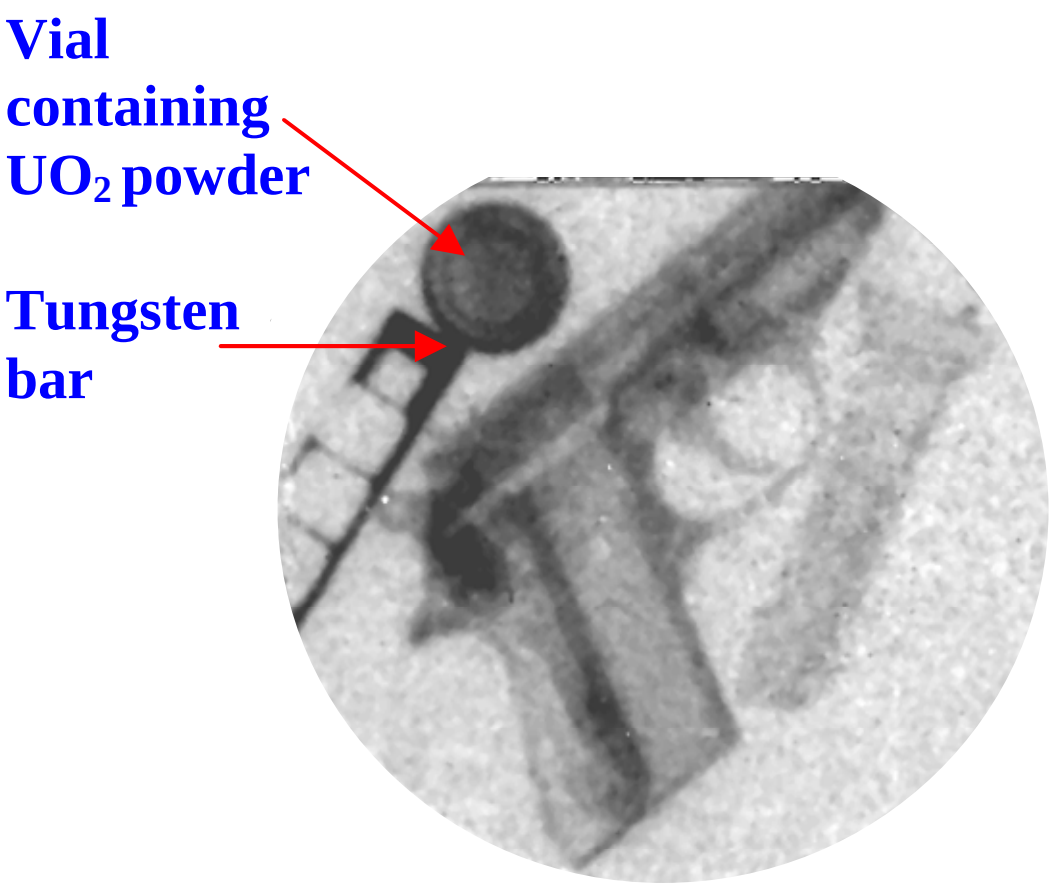

**Fig. 28b Camera 1, $E_n$ = 10.5 MeV**

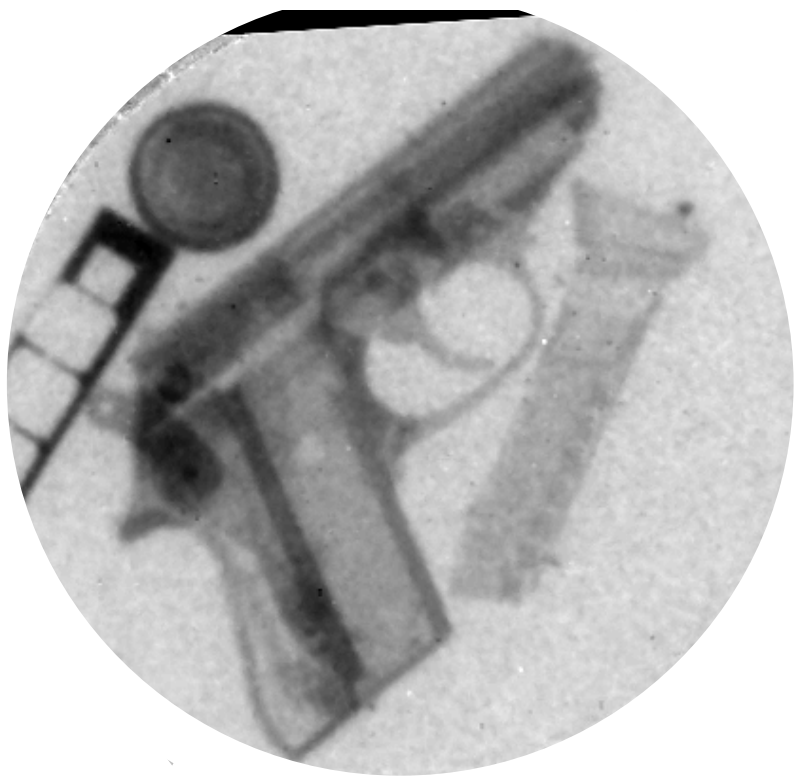

**Fig. 28c Camera 2, $E_n$ = 7.3 MeV**

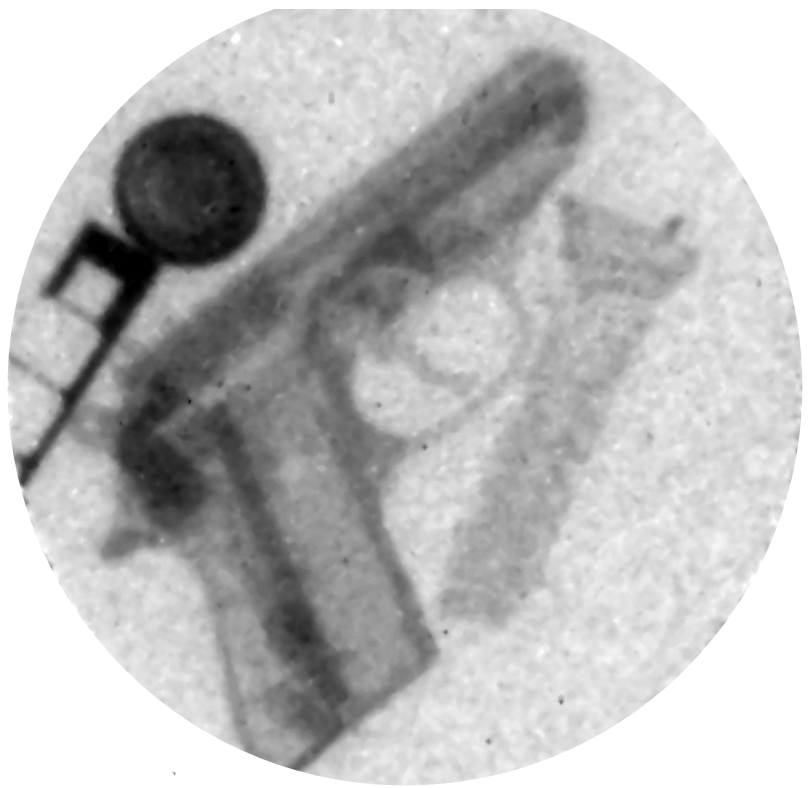

**Fig. 28d Camera 3, $E_n$ = 3.1 MeV**

## 2.3 Detection efficiency

A detailed calculation of the distribution of energy deposited in a scintillating fiber core has been performed using the GEANT 3.21 code [40]. The simulated setup consisted of a 200×200×30 $mm^3$ fiber screen, uniformly irradiated at 5 different neutron energies (2, 4, 7.5, 10 and 14 MeV) by a 200×200 $mm^2$ mono-energetic parallel-beam of neutrons impinging on the screen face. Fig. 29 shows a schematic configuration of 9 fibers located in the center of the screen and a magnified view of the central fiber (0.5×0.5 $mm^2$ polystyrene core, 20 μm thick PMMA cladding and 16 μm thick EMA paint).

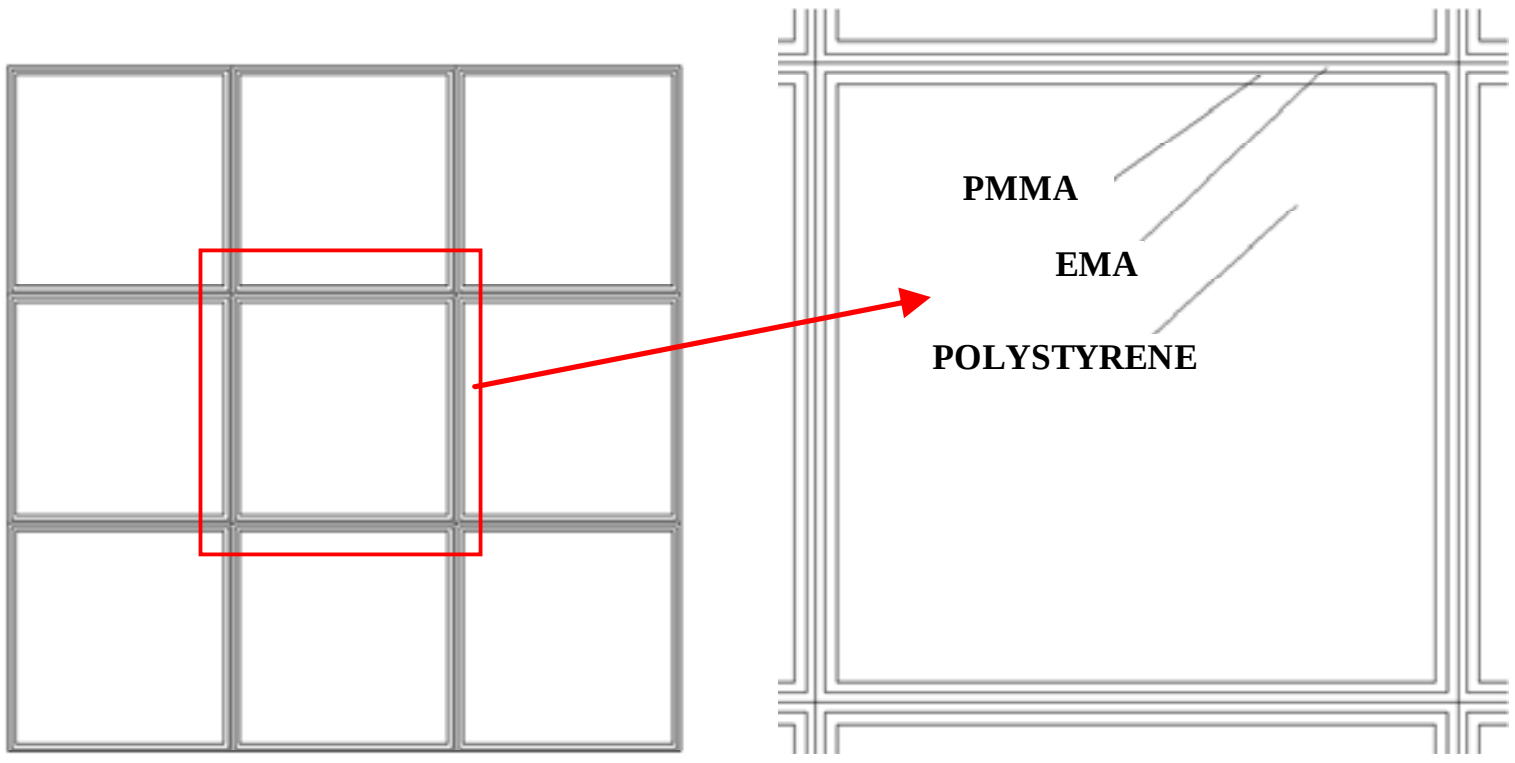

**Fig. 29 Schematic of 9-fiber array (left) and magnified view of central fiber (right)**

The simulation calculates the energy deposited by protons in the core of the central fiber. In the simulation, the same number of neutrons ($6\times10^9$) uniformly incident on the 200×200 $mm^2$ area was employed at each of the 5 neutron energies.

This simulation permitted calculating the efficiency for creating light within the fiber scintillator screen. Calculation of neutron detection efficiencies performed with the above simulation setup agree with analytical calculations based on evaluated neutron cross-section databases such as ENDF/B-VII [41].

Based on simulated and measured responses of each of TRION's components, a custom Monte-Carlo program developed using the Matlab [37] environment was used to calculate the average number of CCD electrons created per incident neutron. Table 1 shows the results of these calculations.

**Table 1 TRION detection efficiencies**

| Neutron Energy (MeV) | Neutron Detection Efficiency % | Number of scintillation photons created per detected event | Number of CCD electrons created per incident neutron |
|---|---|---|---|
| 2 | 30 | 721 | 1015 |
| 4 | 21 | 2260 | 2085 |
| 7.5 | 14 | 5424 | 3347 |
| 10 | 11 | 8258 | 4136 |
| 14 | 8 | 10425 | 4045 |

As can be seen in this table and in Fig. 30, the number of CCD electrons created per incident neutron increases with energy up to 10 MeV and then starts to decline moderately. This can be attributed to the fact that, at the higher neutron energies, the increase in number of scintillation photons created per detected event does not offset the decrease in neutron detection efficiency.

A more complete description of signal analysis in TRION can be found in ref. [26].

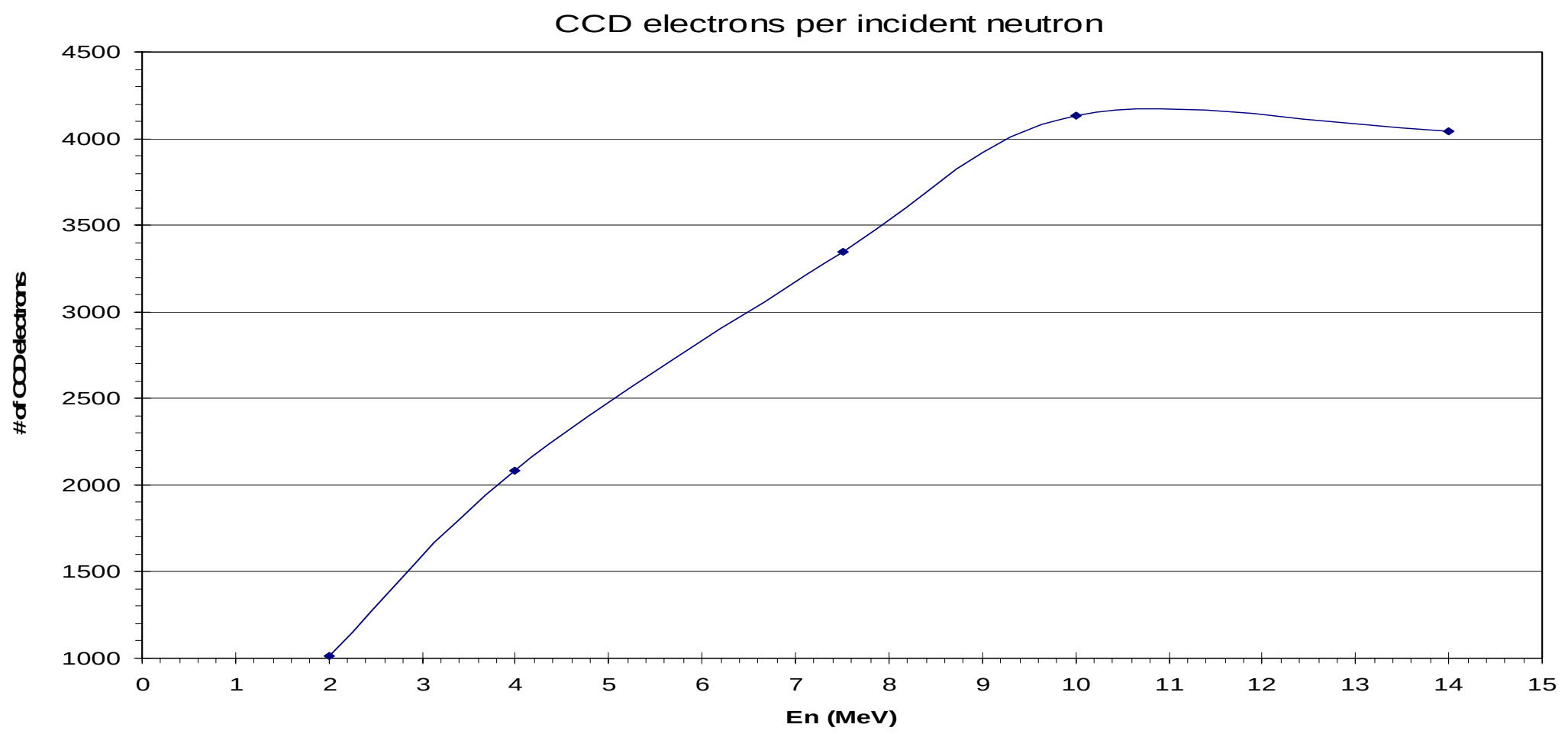

**Fig. 30 Number of CCD-electrons created per incident neutron vs. neutron energy**

## 3. Summary

In this paper we have described the concept, design and performance of TRION, a novel integrative fast-neutron imaging detector enabling energy selection using neutron-TOF. This detector is well-suited for Pulsed Fast-Neutron Transmission Spectroscopy (PFNTS), a promising technique for an automated explosives detection system. In terms of spatial resolution and event-rate handling capability, TRION represents significant progress compared to existing counting systems, along the way to a viable and cost-effective detector for operational PFNTS inspection systems.

We have presented here the evolution from a single time-frame to a quad-time-frame system. At present, the development of an octo-time-frame system is in progress. It employs the concept of an image splitter which will require the use of only a single high-resolution CCD camera.

As demonstrated above, "conventional" visual inspection is also feasible with TRION due to its high spatial resolution and high contrast for low-Z materials, even if these are shielded by high-Z substances.

A crucial component of these multi-frame systems is a large area, nanosecond-fast optical preamplifier that enables high optical light detection efficiency with multi-frame independently-gated intensified cameras.

Acknowledgments

We wish to thank A. Schoen and E. Cheifetz from El-Mul Technologies and I. Cox from Photek Ltd. for their support and fruitful discussions.

The work on TRION Gen.2 was supported by the U.S. Transportation Security Laboratory (TSL) under: HSTS04-05-R-RED108.